\documentclass[fleqn,usenatbib]{mnras}

\usepackage{newtxtext,newtxmath}

\usepackage[T1]{fontenc}

\DeclareRobustCommand{\VAN}[3]{#2}
\let\VANthebibliography\thebibliography
\def\thebibliography{\DeclareRobustCommand{\VAN}[3]{##3}\VANthebibliography}

\usepackage[T1]{fontenc}
\usepackage{amsmath}
\usepackage{bm}
\usepackage{comment}
\usepackage{xcolor}
\usepackage{graphicx}	
\usepackage{natbib}

\title[Dust Concentration in Non-Ideal MHD Discs]{Dust Concentration Via Coupled Vertical Settling and Radial Migration in Substructured Non-Ideal MHD Discs and Early Planet Formation}

\author[C.-Y. Hsu et al.]{
Chun-Yen Hsu,$^{1,2}$\thanks{E-mail: kdj8qp@virginia.edu}
Zhi-Yun Li,$^{1,2}$
Yisheng Tu,$^{1,2}$
Xiao Hu,$^{1,3}$
Min-Kai Lin,$^{4,5}$
\\
$^{1}$Department of Astronomy, University of Virginia, Charlottesville, VA 22904, USA\\
$^{2}$ Virginia Institute of Theoretical Astronomy, University of Virginia, Charlottesville, VA 22904, USA\\
$^{3}$Department of Astronomy, University of Florida, Gainesville, FL 32608, USA\\
$^{4}$Institute of Astronomy and Astrophysics, Academia Sinica, Taipei 10617, Taiwan\\
$^{5}$Physics Division, National Center for Theoretical Sciences, Taipei 10617, Taiwan
}

\date{Accepted XXX. Received YYY; in original form ZZZ}

\pubyear{\the\year{}}

\begin{document}
\label{firstpage}
\pagerange{\pageref{firstpage}--\pageref{lastpage}}
\maketitle

\begin{abstract}

We investigate the dynamics of dust concentration in actively accreting, substructured, non-ideal MHD wind-launching disks using 2D and 3D simulations incorporating pressureless dust fluids of various grain sizes and their aerodynamic feedback on gas dynamics. Our results reveal that mm/cm-sized grains are preferentially concentrated within the inner 5–10 au of the disk, where the dust-to-gas surface density ratio (local metalicity $Z$) significantly exceeds the canonical 0.01, reaching values up to 0.25. This enhancement arises from the interplay of dust settling and complex gas flows in the meridional plane, including midplane accretion streams at early times, midplane expansion driven by magnetically braked surface accretion at later times, and vigorous meridional circulation in spontaneously formed gas rings. The resulting size-dependent dust distribution has a strong spatial variation, with large grains preferentially accumulating in dense rings, particularly in the inner disk, while being depleted in low-density gas gaps. In 3D, these rings and gaps are unstable to Rossby wave instability (RWI), generating arc-shaped vortices that stand out more prominently than their gas counterparts in the inner disk because of preferential dust concentration at small radii. The substantial local enhancement of the dust relative to the gas could promote planetesimal formation via streaming instability, potentially aided by the ``azimuthal drift'' streaming instability (AdSI) that operates efficiently in accreting disks and a lower Toomre $Q$ expected in younger disks. Our findings suggest that actively accreting young disks may provide favorable conditions for early planetesimal formation, which warrants further investigation.
\end{abstract}




\section{Introduction}
\label{sec:Introduction}

Rings and gaps are the most common substructures in high-angular-resolution dust continuum \citep[e.g.,][]{ALMA15, Andrews18} and molecular line emission \citep[e.g.,][]{Oberg21} and in protoplanetary disks. Although these axisymmetric substructures are frequently observed, non-axisymmetric features such as spiral arms, lobes, arcs, vortices, and crescents are also detectable in dust continuum \citep[e.g.,][]{Grady13,Dong18,Boehler21, Marr22} and high-resolution molecular line observations \citep[e.g.,][]{Teague19,Pinte19,Oberg21}. 

These non-axisymmetric substructures are thought to be generated by the Rossby Wave Instability (RWI) \citep[e.g.,][]{Lovelace99, Li2000}. The RWI is triggered when a local Rossby wave becomes trapped in a steep density or pressure bump within a non-self-gravitating disk. Such bumps are believed to form at the edges of planet-opened gaps \citep[][]{Huang19,Wu24}, dead zone boundaries \citep[][]{Miranda17}, and in rings and gaps spontaneously produced in wind-launching non-ideal MHD disks \citep[][]{Suriano18,Suriano19,Hu22,Nolan23,Hsu24}. The primary goal of this paper is to investigate the extent to which dust can be concentrated in the spontaneously produced gas substructures in actively accreting, magnetic wind-launching disks through global 2D (axisymmetric) and 3D non-ideal MHD simulations with coupled dust and gas dynamics. 

Dust concentration is essential to study because it provides conditions that are conducive to planetesimal formation through, e.g., the streaming instability \citep[SI;][]{Youdin05,Johansen07}, a leading mechanism to enhance the dust-to-gas ratio to the point of gravitational collapse, forming planetesimals \citep[]{Johansen09_2,Simon17}. The SI is typically studied with high-resolution local shearing box simulations \citep[e.g.][]{Johansen14,Lesur22}. A crucial quantity in the vertically stratified version of such simulations that determines whether SI-induced dust clumping can lead to gravitational collapse and planetesimal formation is the dust-to-gas surface density ratio (often referred to as metalicity $Z$). For example, \cite{Rixin21} found critical values for $Z$ for strong clumping of $\sim 0.007$ (or lower) for relatively large grains with Stokes number ${\rm St} \gtrsim 10^{-2}$. As ${\rm St}$ decreases from $\sim 10^{-2}$ to $\sim 10^{-3}$, the critical value increases from $\sim 0.02$ to $\sim 0.04$, higher than the canonical value of $0.01$. Whether the dust-to-gas surface density ratio $Z$ can be significantly increased above 0.01 in localized regions is therefore an important question to address.

The most widely discussed scenario for concentrating dust radially relative to gas (and thus increasing the local $Z$) is through pressure maxima \citep[as reviewed, e.g., by][]{Birnstiel24}, where relatively large grains drift toward the maxima and get trapped there. However, this mechanism works well only when the gas is rather quiescent, with a very low level of turbulence, as may be expected in a late stage of disk evolution (e.g., Class II), when the mass accretion rate is low, at least near the midplane to which large dust tends to settle. Whether dust can be concentrated preferentially over gas in more actively accreting younger (Class 0 and I) disks to facilitate earlier planet formation is less well explored. 

One way to radially concentrate the dust relative to the gas in an actively accreting disk is to have a vertically varying accretion flow and have the gas and dust experiencing different parts of that flow. This possibility was recently demonstrated by \cite{Okuzumi25} using a semi-analytic model where the disk surface layer is assumed to accrete faster than the midplane region. A motivation for the preferential surface accretion comes from non-ideal MHD simulations of magnetic wind-launching disks \citep[e.g.][]{Bethune17,Suriano17,Suriano18,Riols19,Cui21,Iwasaki24}, where the disk surface is expected to accrete faster because it is more ionized than the midplane region and thus better coupled to the magnetic field. However, the actual disk structure and flow pattern in non-ideal MHD wind-launching disks are quite complex, including spontaneous formation of substructures (rings and gaps), spatially and temporally variable fast accreting streams near both the disk surface and midplane, as well as vigorous meridional circulations \citep[e.g.][]{Hu22} and vortices \citep[e.g.][]{Hsu24}. This paper aims to determine whether significant radial dust concentration with locally enhanced $Z$ can occur in such a highly dynamic environment.   

The paper is organized as follows. Section \S \ref{sec:simulation} introduces the governing equations, simulation domain, boundary conditions, initial conditions, and the chemical network for determining the non-ideal MHD coefficients used in our simulations.   
Section \S \ref{sec: results} presents the simulation results, focusing on vertical settling and radial migration of the dust that preferentially enhance the dust over the gas in localized regions of the disk. We find, in particular, large local enhancements in the dust-to-gas surface density ratio $Z$ over the canonical value of 0.01 in the 5-10~au region. In Section \S \ref{sec: discussion}, we discuss the implications of dust concentration on streaming instability and potential early planetesimal formation in actively accreting disks. Our main conclusions are summarized in \S \ref{sec: conclusion}.

\section{Simulation} 
\label{sec:simulation}

We use Athena++ \citep{stone20} with a pressureless dust fluid module  \citep{Huang22,Huang25}  to solve the non-ideal MHD gas and dust fluid equations:
\begin{equation}
  \frac{\partial \rho_{\rm g}}{\partial t} + \nabla \cdot (\rho_{\rm g} {\rm \bf V}_{\rm g}) = 0, \label{eq:gas_continuity}
\end{equation}
\begin{equation}
  \frac{\partial (\rho_{\rm g} {\rm \bf V}_{\rm g})}{\partial t} + \nabla \cdot (\rho_{\rm g} {\rm \bf V}_{\rm g} {\rm \bf V}_{\rm g} + {\rm P^*} {\rm {\bf I}}  - \frac{{\rm \bf B} {\rm \bf B} }{4 \pi}) = - \rho_{\rm g} \nabla \Phi + \rho_{\rm d} \frac{{\rm \bf V}_{\rm d} - {\rm \bf V}_{\rm g}}{\tau_{\rm s}}, \label{eq:gas_momentum}
\end{equation}
\begin{align}
  \frac{\partial E_{\rm g}}{\partial t} + \nabla \cdot  \left[(E_{\rm g} + {\rm P^*}){\rm \bf V}_{\rm g} - \frac{{\rm \bf B} ({\rm \bf B} \cdot {\rm \bf V}_{\rm g})}{4 \pi} + \frac{1}{c} \left( \eta_O {\rm \bf J} + \eta_{\rm AD} {\rm \bf J}_{\perp} \right) \times {\rm \bf B}  \right] \notag \\
  = - \rho_{\rm g} ({\rm \bf V}_{\rm g} \cdot \nabla \Phi) - \Lambda_{\rm c} + \rho_{\rm d} \frac{{\rm \bf V}_{\rm d} - {\rm \bf V}_{\rm g}}{\tau_{\rm s}} \cdot {\rm \bf V}_{\rm g} + \omega_{\rm d} \rho_{\rm d} \frac{|{\rm \bf V}_{\rm d} - {\rm \bf V}_{\rm g}|^2}{\tau_{\rm s}}, \label{eq:gas_energy}
\end{align}
\begin{equation}
\frac{\partial \rho_{\rm d}}{\partial t} + \nabla \cdot (\rho_{\rm d} {\rm \bf V}_{\rm d} ) = 0, \label{eq:dust_continuity}
\end{equation}
\begin{equation}
\frac{\partial \rho_{\rm d} ({\rm \bf V}_{\rm d})}{\partial t} + \nabla \cdot (\rho_{\rm d} {\rm \bf V}_{\rm d} {\rm \bf V}_{\rm d}) = - \rho_{\rm d} \nabla \Phi + \rho_{\rm d} \frac{{\rm \bf V}_{\rm g} - {\rm \bf V}_{\rm d}}{\tau_{\rm s}}, \label{eq:dust_momentum}
\end{equation}
and the induction equation
\begin{equation}
  \frac{\partial {\rm \bf B}}{\partial t} = \nabla \times ({\rm \bf V}_{\rm g} \times {\rm \bf B}) - \frac{4 \pi}{c} \nabla \times \left[ \eta_O {\rm \bf J} + \eta_{\rm AD} {\rm \bf J}_{\perp} \right] , \label{eq:induction}
\end{equation}
where $\rho_{\rm g}$ and ${\rm \bf V}_{\rm g}$ are gas mass density and velocity, $\rho_{\rm d}$ and ${\rm \bf V}_{\rm d}$ are dust mass density and velocity, ${\rm \bf B}$ is the magnetic field, ${\rm P^*} = P_{\rm g} + B^2/(8\pi)$ is the total (thermal [$P_{\rm g}$] and magnetic)  pressure, ${\rm \bf I}$ is the identity tensor, $\Phi = -GM/r$ is the gravitational potential of the central star, $E_{\rm g} = \rho_{\rm g} V_{\rm g}^2/2 + {P_{\rm g}}/(\gamma - 1) + B^2/(8 \pi)$ is the energy density, $\gamma$ is the adiabatic index, ${\rm \bf J}$ is the current density, ${\rm \bf J}_{\perp} = {\rm \bf B} \times ({\rm \bf J} \times {\rm \bf B}) / (B^2)$ is the component of ${\rm \bf J}$ perpendicular to the magnetic field, $\eta_O$ and $\eta_{\rm AD}$ are the Ohmic and ambipolar diffusivities, and $\Lambda_{\rm c}$ is the cooling term.
The aerodynamic drag between gas and dust is included in the last terms of Eq. \ref{eq:gas_momentum} and \ref{eq:dust_momentum}. The dust fluid module in Athena++ assumes linear drag law, so the stopping time $\tau_{\rm s}$ is independent of velocity.
The last two source terms on the right hand side of the gas energy equation (Eq.~\ref{eq:gas_energy}) are drag and frictional heating terms, respectively. The parameter $\omega_{\rm d}$ is used to control the level of frictional heating, and we set $\omega_{\rm d} = 1$ to assume all the dissipation is deposited to the gas.

\subsection{Simulation domain} \label{subsection:simulation_domain}

The simulation domain is the same as the \citet{Hsu24}. We perform 2D (axisymmetric) and 3D simulations using spherical coordinates ($r, \theta, \phi$), where the quantities in 2D are independent of $\phi$. The simulation domain spans 1 to 316~au radially, 0.05 to $\pi$-0.05 in the polar direction, and, for 3D simulations, 0 to $2\pi$ azimuthally. The radial grid follows a logarithmic spacing with 80 base cells and an adjacent cell size ratio of 1.07416. The polar grid is uniform with 96 base cells, and 3D simulations use 32 base cells in the azimuthal direction. We apply three levels of static mesh refinement (SMR), with each level refining the grid by a factor of 2. The finest level extends from 10 to 100 au radially and spans approximately 2.5 scale heights (~0.13 radians) above and below the mid-plane, resolving the disk scale height with ~12.6 grid cells. The second and third refinement levels cover, respectively, 5.57–176~au and 3.14–316~au radially,  and -0.25 to +0.25 radians and -0.5 to +0.5 radians of the midplane in the polar direction. 

\subsection{Boundary conditions} 
\label{subsection:boundary conditions}

We implemented modified outflow boundary conditions at both the inner and outer radial boundaries. At the inner boundary, instead of copying the gas density and pressure from the innermost active zone to the ghost zones, we extended these quantities using the same power-law profile applied during gas initialization. For gas velocity, the azimuthal component follows a Keplerian profile, while the radial and polar components are copied from the innermost active zone, ensuring that no external mass enters the simulation domain. Reflective boundary conditions are applied in the polar ($\theta$) direction, and periodic boundary conditions are used in the azimuthal direction.

\subsection{Initial conditions} \label{initial_conditions_subsection}

We perform two types of simulations: (i) 2D simulations initialized in hydrostatic equilibrium, with a magnetic field derived from the vector potential to ensure a divergence-free $B$ field (i.e. $\nabla \cdot {\rm \bf B} = 0$); and (ii) a 3D simulation that shares the same initial conditions as the 2D simulations, with an initially axisymmetric setup in the azimuthal direction.

 The initial conditions of the gas and the magnetic field are the same as in \cite{Hsu24}, and the initial temperature and gas density profiles follow those used in \cite{Hu22}. Specifically, we divide the simulation domain into a cold and dense disk and a hot low-density corona, maintaining a constant aspect ratio of $h/r=0.05$,  where $h$ is the height of the disk scale. The cold and dense disk is confined within two scale heights above and below the midplane, defined as $\pi/2 - \theta_0 <\theta < \pi/2 + \theta_0$, where $\theta_0=\arctan{(2h/r)}$. The gas density and midplane temperature both follow power-law distributions with indices $p=-1.5$ and $q=-1$, respectively:
\begin{eqnarray}
\rho(r,\pi/2) = \rho_0(r/r_0)^{p},  \\ \nonumber 
T(r,\pi/2) = T_0(r/r_0)^{q} 
\label{eq:profile}
\end{eqnarray}
where $r_0=1$~au is the radius of the inner boundary of the computational domain, and $\rho_0=2.667 \times 10^{-10} {\rm g/cm^{-3}}$ and $T_0=570$~K are the density and temperature at $r_0$. To ensure a smooth transition between the cold disk and hot corona, we adopt the following vertical profile for the temperature:  
\begin{equation}
T(r,\theta)=
\begin{cases}
T(r,\pi/2) & \text{if }  |\theta-\pi/2| < \theta_0 \\
T(r,\pi/2)\ {\rm exp}[(|\theta-\pi/2|\\
\ -\theta_0)/\theta_0 \times\ln(160)]
  & \text{if } \theta_0 \leq |\theta-\pi/2| \leq 2\theta_0  \\
160\ T(r,\pi/2); & \text{if } |\theta-\pi/2| > 2\theta_0\\
\end{cases}\label{eq:T}
\end{equation}
We use a quick $\beta_{\rm cool}$ cooling scheme with a cooling timescale of only $10^{-10}$ of the local orbital period, so the temperature profile is effectively fixed over time.
The vertical density profile is generated based on hydrostatic equilibrium, i.e.,$v_r=v_\theta=0$. 

The initial magnetic field is computed from the magnetic vector potential used in \cite{Zanni07}:
\begin{align}
 & B_r (r, \theta) = \frac{1}{r^2 {\rm sin}\theta} \frac{\partial A_\phi}{\partial\theta}, \label{eq:Br} \\
 & B_\theta (r, \theta) = -\frac{1}{r\ {\rm sin}\theta} \frac{\partial A_\phi}{\partial r}, \label{eq:Btheta} \\
 & B_\phi = 0, \label{eq:Bphi}
\end{align}
with
\begin{align}
 & A_\phi (r, \theta) = \frac{4}{3}r_0^2 B_{\rm p,0} \left( \frac{r {\rm sin}\theta}{r_0}\right)^{\frac{3}{4}} \frac{1}{(1 + 2{\rm cot^2 \theta})^{5/8}}
\end{align}
where $B_{\rm p,0}$ sets the scale for the poloidal field strength. The magnetic field setup is the same as that of \cite{Bai17}, \cite{Suriano18}, and \cite{Hu22}. In all our simulations, $B_{\rm p,0}$ is set by plasma $\beta = 10^3$ in the mid-plane. 

We set the initial total dust mass density $\rho_{\rm d}$ to $1\%$ of the initial gas mass density $\rho_{\rm g}$ and the initial dust velocity the same as that of the gas. In the relatively dense regions within four gas scale heights of the midplane, we compute the dust stopping time based on the dust size $a$:
\begin{align}
 &\tau_{\rm s} = \frac{\rho_{\rm m}}{\rho_{\rm g}} \frac{a}{v_{\rm th}}
\end{align}
where $\rho_{\rm m}$ is the grain material density, taken to be 3~${\rm g/cm^{3}}$ here and in the chemistry network (\S \ref{sec: chemistry and non-ideal}). $v_{\rm th}$ is the thermal speed of the gas. In the low-density regions above and below four gas scale heights from the midplane, we fix the Stoke number ${\rm St}=\tau_{\rm s}\Omega_{\rm K}$ at $0.01$ to make the simulation run more stably. This assumption does not significantly affect our conclusions since they will be based primarily on grains that settle well below four gas scale heights.  

\subsection{Chemical network and the non-ideal MHD coefficients} 
\label{sec: chemistry and non-ideal}

The chemical network and the non-ideal MHD coefficients used in this work are the same as in \cite{Hsu24} (see their Sec. 3 and Appendix A for details). Briefly, we used a simplified chemical network following \cite{Umebayashi90}, \cite{Nishi91}, and \cite{grassi19}, which include the element H, He, C, O, and Mg, an MRN-type \citep{Mathis77}  power-law grain size distribution (with $a_{\rm min} = 0.5$ $\mu$m and $a_{\rm max} = 25$ $\mu$m, and a power-law index of $-3.5$), and gas-phase, gas-grain, and grain-grain reactions. Once the charge abundances are computed from the chemical network, they are stored in a lookup table and referenced during simulations to calculate the Ohmic and ambipolar diffusivities for each computational cell using the standard formulae (see, e.g., Eq.~[18]-[24] of \citealt{Hsu24}).  

To mimic the expected better magnetic coupling (and thus reduced diffusivity) due to higher ionization levels expected in the lower density regions near the disk surface and in the disk wind, we follow \cite{Suriano18} and multiply the computed Ohmic and ambipolar diffusivities by the following $\theta$ dependence: 
\begin{align}
 & f(\theta) = 
\begin{cases} 
{\rm exp} \left(-\frac{{\rm cos}^2(\theta + \theta_0)}{2(h/r)^2} \right) & \mbox{if} \quad  \theta < \frac{\pi}{2} -  \theta_0 \\
1 & \mbox{if} \quad \frac{\pi}{2} - \theta_0 \leq \theta   \leq \frac{\pi}{2} + \theta_0 \\
{\rm exp} \left(-\frac{{\rm cos}^2(\theta - \theta_0)}{2(h/r)^2} \right) & \mbox{if} \quad \theta > \frac{\pi}{2} + \theta_0 
\end{cases}. \label{eq:theta_depend_non_ideal} 
\end{align}

\section{Dust concentration in substructured accreting disks} 
\label{sec: results}

Table. \ref{table_cases} summarizes the models used in this study, including both 2D and 3D simulations. 
The 2D models include model 2D1mm with a single dust fluid with 1~mm grain size, model 2D10mm01mm5bins with 5 dust bins from $a_{\rm min}=0.1$ to $a_{\rm max}=10$~mm, and model 2D10mm10um30bins with 30 dust bins from $a_{\rm min}=0.01$~mm to $a_{\rm max}=10$~mm. We label them based on their minimum and maximum grain sizes and the number of dust bins equally spaced logarithmically.
A 3D model with single-sized (1~mm) grains is also included in this work. All models assume an initial total dust-to-gas mass ratio of $\epsilon=0.01$ within the simulation domain. Dust feedback (drag forces on both Eq. \ref{eq:gas_momentum} and Eq. \ref{eq:dust_momentum}) is included in all models except for Model 2D1mm$\_$nofeedback, which serves as a control for our discussion of the effects of the dust aerodynamic feedback on the gas in \S~\ref{subsec: dust_feedback}. 

\begin{table*}
\caption {
Models
\label{table_cases}
} 
\begin{tabular}{lllll}
  Name  & Dust fluid number  & Initial $\epsilon$ & size distribution & dust feedback \\
        \hline
    2D1mm   &  1  & 0.01 & single & yes \\
    2D10mm01mm5bins  & 5    & 0.01 & MRN & yes \\
    2D10mm10$\mu$m30bins  & 30    & 0.01 & MRN & yes \\
    2D1mm$\_$nofeedback & 1 & 0.01 & single & no \\
    3D1mm     & 1   & 0.01 & single & yes 
\end{tabular}
\end{table*}

\subsection{Overview of gas substructure and dust concentration} \label{subsec: 2D_single_fluid}

\begin{figure*}
    \centering
    \includegraphics[width=\linewidth]{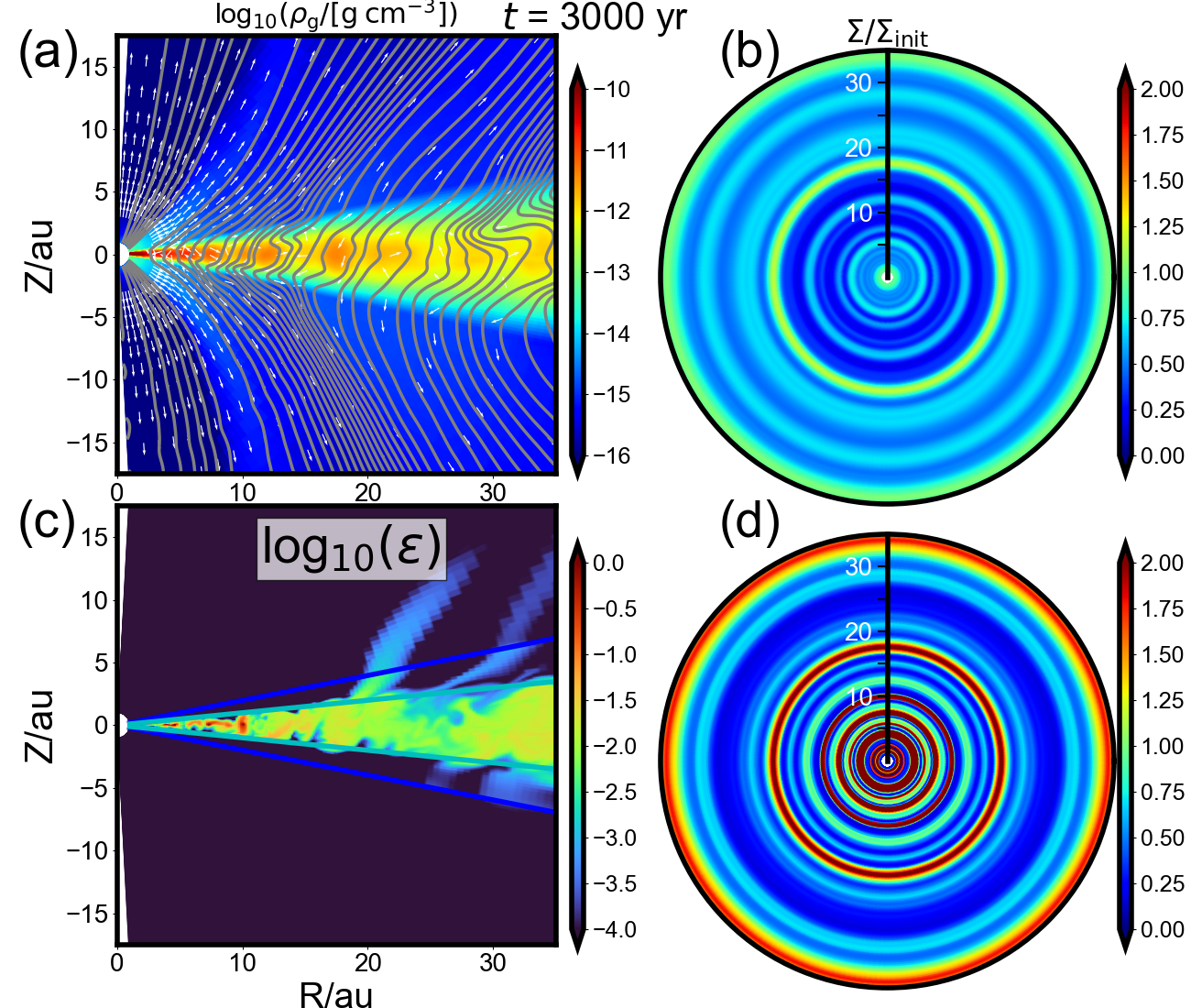}
    \caption{Gas and dust substructures in a wind-launching disk (Model 2D1mm)  at a representative time $t$ = 3000~yr up to 35 au in radius. Plotted in panel (a) are the gas mass density distribution (color map), velocity vectors (white arrows), and magnetic field lines (gray lines) on a meridional plane. Panel (c) is the dust-to-gas mass ratio $\epsilon$ on a logarithmic scale, with the straight blue (cyan) lines marking 4 (2) gas scale heights away from the midplane. Panels (b) and (d) display the gas and dust surface densities of the disk normalized to their respective initial value, highlighting the formation of prominent rings and gaps in both gas and dust, with the dust substructures generally having higher contrast. An animated version of the figure can be found at \url{https://figshare.com/s/e5fd46539aee9cd263f8}. 
    }
\label{fig:2D_1mm_rings_gaps}
\end{figure*}

\begin{figure*}
    \centering
    \includegraphics[width=\linewidth]{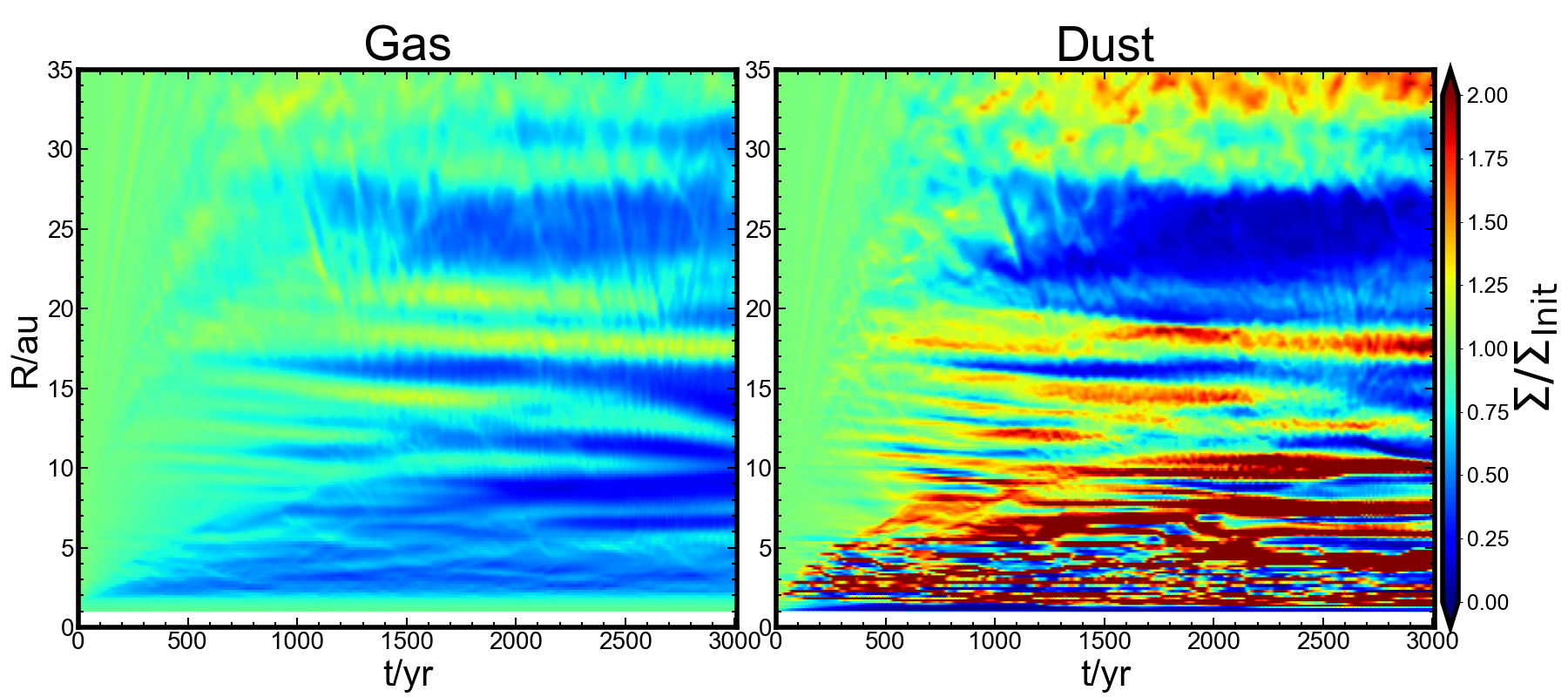}
    \caption{ Time evolution of the surface density distribution of (a) gas and (b) dust of Model 2D1mm as a function of radius, highlighting the formation of prominent rings and gaps in both gas and dust, with the dust substructures generally having higher contrast, as in Fig. \ref{fig:2D_1mm_rings_gaps}. 
    }
\label{fig:2D_1mm_time_evolve}
\end{figure*}
To illustrate the broad features of the gas and dust in the coupled disk and wind system, we start with a simple model 2D1mm that is axisymmetric and has one representative grain size (1~mm).
In Fig. \ref{fig:2D_1mm_rings_gaps}, we show the gas and dust structure in a meridional (left panels) and face-on (right panels) view of the model at a representative time $t = 3000$~yr,  when stable rings and gaps are formed. We used cylindrical coordinates (R, $\phi$, z) in the meridional plots. 
Prominent rings and gaps are formed in both the gas and the dust fluid, adding weight to the formation of rings and gaps in wind-launching disks studied in the literature \citep[e.g.,][]{Suriano18,Suriano19,Riols20,Cui21,Nolan23}.  
In particular, there is a strong variation in the spatial distribution of the magnetic flux, which tends to concentrate more strongly in the gaps (see, e.g.,  Fig.~\ref{fig:2D_1mm_rings_gaps}a)\footnote{We note that the fast cooling adopted in our simulation can, in principle, trigger the Vertical Shear Instability (VSI). However, the relatively strong magnetic field (with an initial plasma-$\beta$ of $10^3$) and magnetic coupling (with an ambipolar Els\"asser number greater than unity) effectively suppress the development of the VSI \citep{Cui20}.}. Our results are also broadly consistent with the simulations of \cite{Hu22}, which included Lagrange dust particles but did not include dust feedback on the gas dynamics. Clearly, including the aerodynamic drag of the dust on the gas does not suppress the formation of the disk substructure in either the gas or the dust; its effects will be discussed in more detail in \S~\ref{subsec: dust_feedback}. 

The most striking feature of Fig.~\ref{fig:2D_1mm_rings_gaps} is that dust and gas are not concentrated to the same degree. Specifically, the surface density distributions of the gas (panel b) and dust (panel d) show that the dust is depleted more than the gas in the relatively wide gap near $\sim 25$~au relative to their respective initial value. In contrast, the dust is concentrated to a larger extent in the rings, particularly at relatively small radii inside $\sim 20$~au at the representative time shown. For example, the ring at $\sim 10$~au has peak gas and dust surface densities of $\sim 0.39$ and $\sim 10.7$ times their respective initial values, indicating a preferential concentration of the dust relative to the gas in the radial direction. This radial dust concentration is further enhanced by vertical dust settling, which can increase the local dust mass density relative to its initial value by a factor much higher than that of the surface density. For example, the red regions in panel (c) show that the dust-to-gas mass ratio $\epsilon$ can reach values not far from unity in the inner disk (inside $\sim 10$~au), making such regions potentially conducive to streaming instability and planetesimal formation; this possibility will be discussed in detail in \S~\ref{sec: discussion} below. 

To further illustrate the process of dust and gas radial concentration over time, we plot in Fig. \ref{fig:2D_1mm_time_evolve} the time evolution of the normalized gas and dust surface densities of Model 2D1mm as a function of time. At $t = 1000$~yr, the dust surface density around $\sim 23$~au exhibits significant depletion relative to the gas, while the dust in the inner disk shows strong concentration. In contrast, the gas does not display a pronounced ring-gap structure in the 20- 30~au region until around $t = 1200$~yr. This difference in the time evolution between gas and dust highlights the role of vertical settling in enhancing the radial dust concentration prior to the formation of the gas ring and gap substructure. 

To examine the size-dependent behavior of dust radial migration and vertical settling, we will next focus on Model 2D10mm01mm5bins, which includes grains of different sizes. It is the reference model that will be discussed in the greatest detail. 

\subsection{Dust concentration from interplay between size-dependent vertical settling and radial migration} \label{subsec: 2D_muti-fluid}

\begin{figure*}
    \centering
    \includegraphics[width=\linewidth]{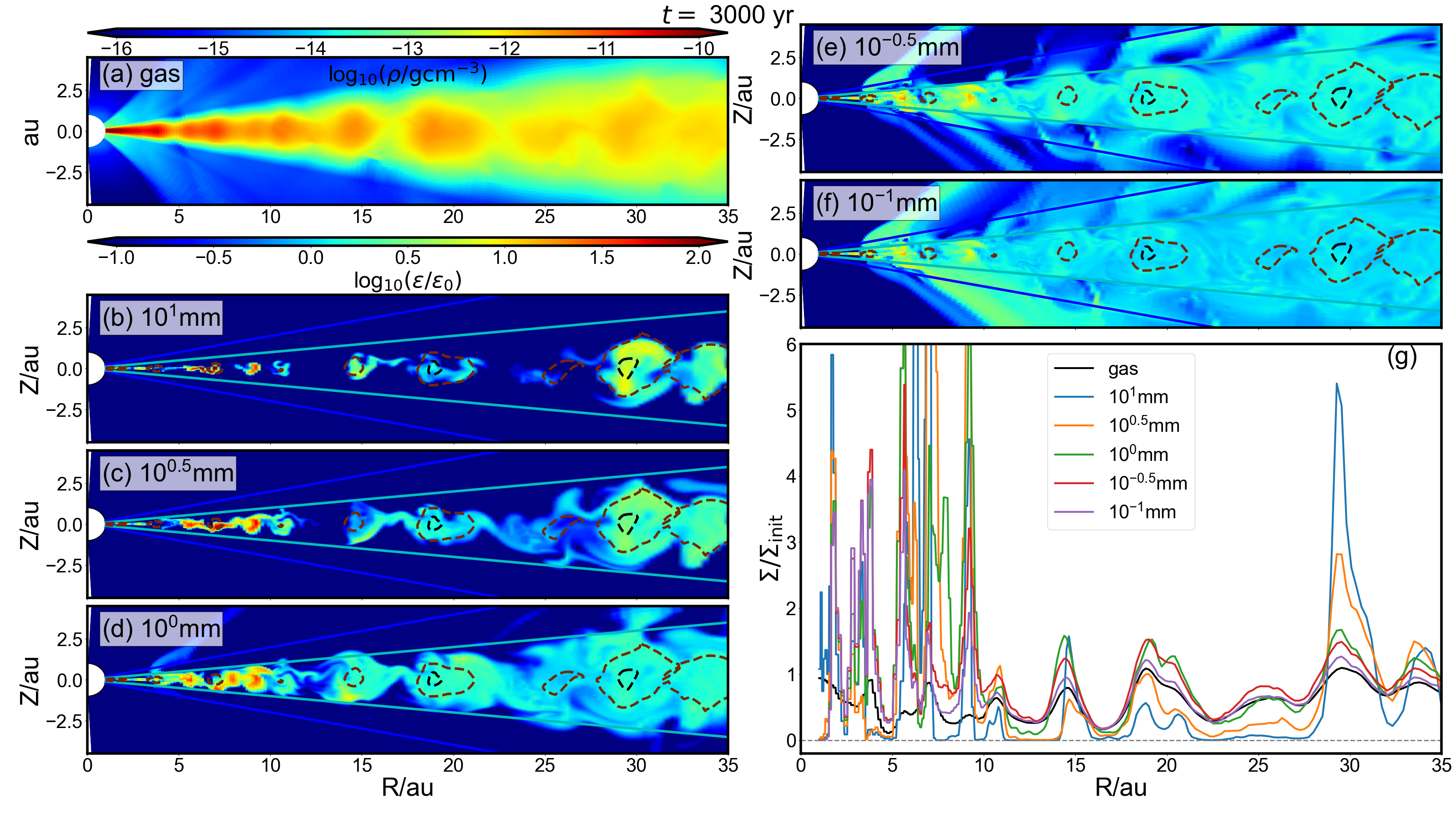}
    \caption{Meridional view of the gas and dust distributions of Model 2D10mm01mm5bins at a representative time $t=3000$~yr. Panel (a) shows the surface density of the gas normalized to its initial value. Panels (b)-(f) are the normalized dust-to-gas mass ratios for grains with sizes of $10^{1}$, $10^{0.5}$, $10^{0}$, $10^{-0.5}$, and $10^{-1}$ mm, respectively. The dotted brown and black contour lines indicate gas mass density levels of 0.6 and 1.0, normalized to their initial values. The cyan and blue lines correspond to two and four scale heights, respectively. The gray dashed lines at 7 and 9~au mark the regions of high dust concentration, as discussed in the main text. An animated version of the figure can be found at \url{https://figshare.com/s/59973715ac7a11c0d7f3?file=52902449}.  
    }
\label{fig: 2D_10mm01mm5bins_rings_gaps}
\end{figure*}

\begin{figure*}
    \centering
    \includegraphics[width=\linewidth]{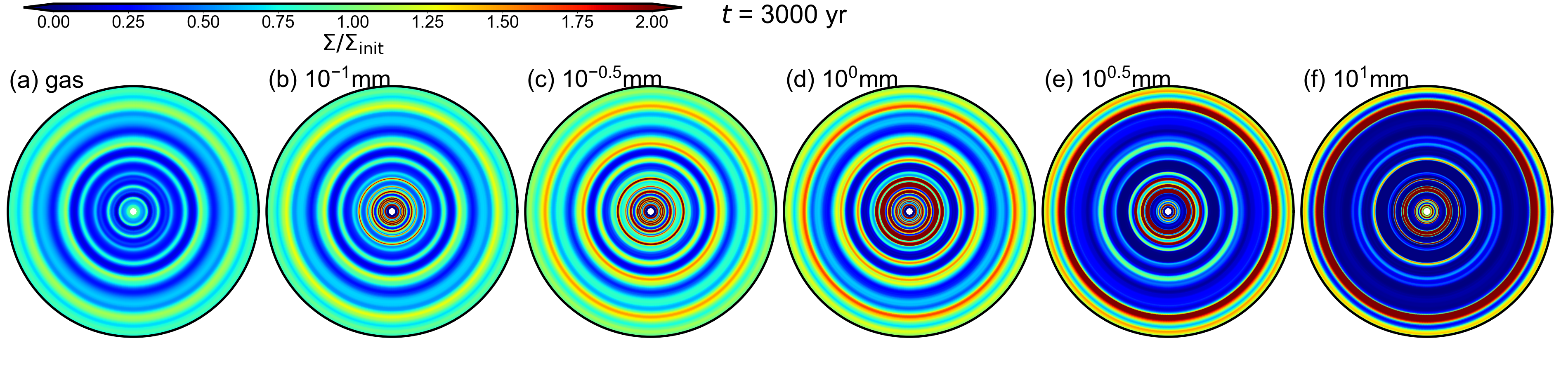}
    \caption{
    The surface densities of the gas and different-sized grains normalized to their initial values within a radius of 35~au, highlighting the formation of prominent rings and gaps in both gas and dust and their differences. An animated version of the figure can be found at \url{https://figshare.com/s/ba4aed22bd3c283c72bb}. 
    }
    \label{fig: 10mm_01mm_5bins_column_density_faceonview}
\end{figure*}

\begin{figure*}
    \centering
    \includegraphics[width=\linewidth]{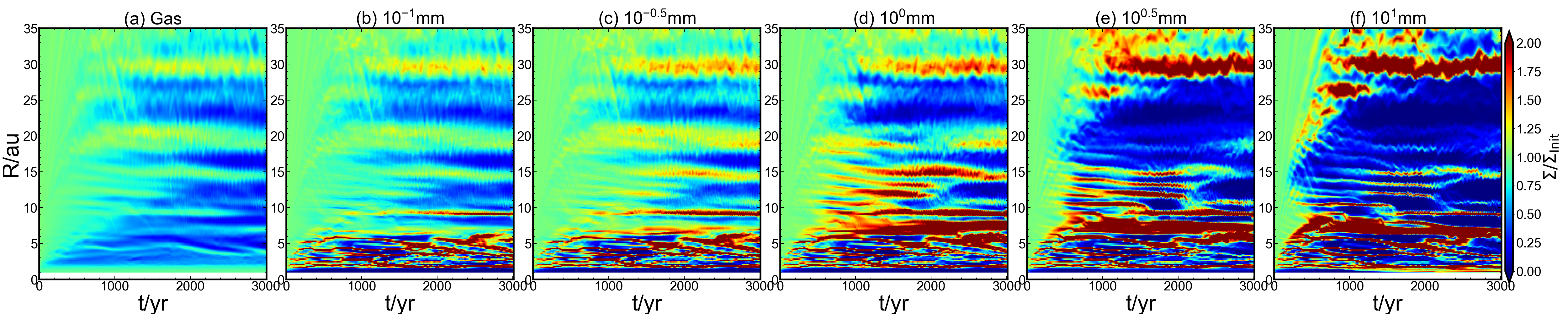}
    \caption{
    Same as Fig. \ref{fig:2D_1mm_time_evolve}, but for Model 2D10mm01mm5bins. 
    }
    \label{fig: 10mm_01mm_5bins_time_evolve}
\end{figure*}

\begin{figure}
    \centering
    \includegraphics[width=1.0\linewidth]
    {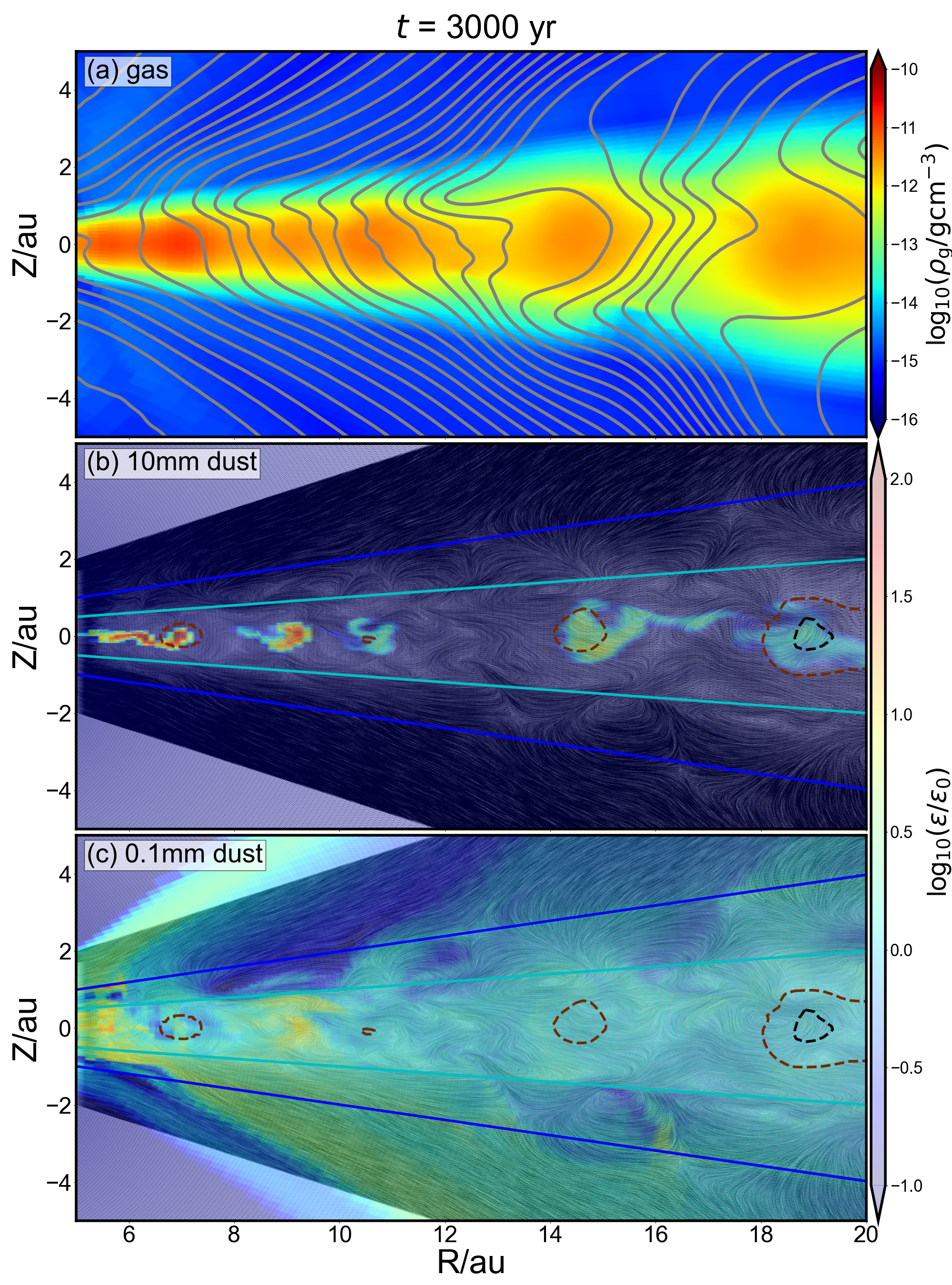}
    \caption{
    Dust concentration and gas velocity field on a meridional plane at the representative time $t=3000$~yr. Plotted in panel (a) are the gas density (color map) superposed with poloidal magnetic field lines (gray lines). Panels (b) and (c) display LIC (line integral convolution) streamlines for poloidal gas motions and the dust-to-gas mass ratios (normalized to their initial values) for the largest and smallest grains, respectively. The dashed lines are iso-contours where the gas density is, respectively, 0.6 (brown) and 1.0 (black) times the initial value, highlighting the gas rings.  Accreting streams and meridional circulations are evident in the LIC plots and especially the animated version of the figure, which can be found at \url{https://figshare.com/s/2d17bd44d82855bc008b}.  
    }
    \label{fig: LIC}
\end{figure}

Model 2D10mm01mm5bins includes grains of five different sizes: $10^{1}$, $10^{0.5}$, $10^{0}$, $10^{-0.5}$, and $10^{-1}$~mm; they satisfy the MRN-type power-law size distribution with a power index of -3.5. The spatial distributions of the gas and the different-sized grains and their time evolution are shown in Figs.~\ref{fig: 2D_10mm01mm5bins_rings_gaps} and \ref{fig: 10mm_01mm_5bins_column_density_faceonview} and their associated animations. Specifically, panel (a) of Fig.~\ref{fig: 2D_10mm01mm5bins_rings_gaps} displays the meridional distribution of the gas mass density, showing prominent rings and gaps at the time shown ($t=3000$~yr). Panels (b)-(f) plot, respectively, the dust-to-gas mass ratio $\epsilon$ for each grain size relative to their initial value. It is immediately clear that different-sized grains are distributed very differently. As expected, larger grains have settled to thinner regions close to the disk midplane at the time shown, since they are pulled by the vertical component of the stellar gravity to move faster toward the midplane relative to the gas. 

A more subtle difference is that the different-sized grains are distributed differently in the radial direction as well. This difference is illustrated in two ways. First, from the normalized dust-to-gas mass ratio plots (panels [b]-[f]), it is clear that while some regions of higher dust concentration (relative to the gas) are located at similar radii for different-sized grains (e.g., the gas ring near $r\sim 9$~au marked by the second vertical dashed line), other regions can have higher concentrations for grains of some sizes but not others. For example, there is a preferential concentration of grains of the three largest sizes ($10$, $10^{0.5}$, and 1~mm; see panels [b-d]) over the gas in the gas ring at $r\sim 7$~au (marked by the first vertical dashed line), but not the two smallest grains ($10^{-0.5}$ and $10^{-1}$~mm). An even more extreme case is the apparent gas gap region between the $\sim 7$ and $\sim 9$~au rings, where the 1~mm grains are preferentially concentrated but not grains of other sizes. 

This example of size-dependent radial dust distribution can be seen more clearly in panel (g), which shows a peak in the dust mass surface density (normalized by its initial value) for 1~mm-sized grains at $\sim 8$~au (see the green curve), but not for the gas or grains of other sizes. There are several additional features worth noting in the panel. We start with the radial variation of the gas surface density distribution relative to its initial value (the black line), which is relatively moderate, with contrasts between the rings and their adjacent gaps typically of order 2. The surface density distributions of the two smallest grains (with sizes of $10^{-1}$ and $10^{-0.5}$~mm) tend to follow the gas beyond a radius of $\sim 10$~au, consistent with the relatively constant values of the dust-to-gas mass ratio for these grain sizes in the outer regions (panel [e-f]). In contrast, the two largest grains (with sizes of $10$ and $10^{0.5}$~mm) are preferentially trapped in the gas ring at $\sim 30$~au relative to the gas, and are severely depleted in the gas gaps between $\sim 10$ and $\sim 25$~au, again consistent with the dust-to-gas mass ratios shown in panels (b) and (c). The situation is quite different inside $\sim 10$~au, where grains of all sizes are strongly concentrated relative to the gas in gas rings (and for at least one dust size, 1~mm, even in the $\sim 8$~au gap, as discussed earlier). This is true even for relatively low-contrast gas rings, such as those at $\sim 9$ and $\sim 5$~au. These column density features are illustrated more pictorially in Fig.~\ref{fig: 10mm_01mm_5bins_column_density_faceonview}, which clearly shows the trapping of the largest grains near $\sim 30$~au, the severe depletion of such grains inside the trap, and the concentration of dust of all sizes at small radii. 

The size-dependence of dust settling, concentration, and radial migration is further illustrated in Fig. \ref{fig: 10mm_01mm_5bins_time_evolve}, where we compare the time evolution of the radial surface density distributions of five different dust sizes alongside the gas.  Compared to the single-size (1~mm) dust case shown in Fig. \ref{fig:2D_1mm_time_evolve}, the $10$~mm dust becomes significantly depleted around $\sim$15~au at a time as early as $t\sim 400$ yr, while the $10^{0.5}$~mm dust does so near $\sim$18~au as early as $t\sim 500$ yr. These results support the idea that larger grains settle toward the mid-plane earlier and migrate inward more efficiently than smaller grains.  Interestingly, we also find that the gas begins to deplete between 20–30~au at $t\sim$800 yr, earlier than what is observed in Model 2D1mm. This suggests that dust feedback can indeed alter the gas dynamics in certain regions, though it does not fundamentally prevent the formation of gas rings and gaps, as we discuss in more detail in \S\ref{subsec: dust_feedback}.  

The size-dependent dust spatial distribution is controlled by the interplay between vertical settling and radial migration, which both depend on grain size. Larger grains tend to vertically settle more quickly towards the midplane and radially migrate more quickly towards the central star. These tendencies are modified by the gas dynamics and substructure. Specifically, the gas in the disk typically has two distinct components -- a fast accretion layer whose location fluctuates around the midplane and meridional circulations above and below the accretion layer, especially at relatively early times and at relatively large radii, as illustrated in panels (b) and (c) of Fig.~\ref{fig: LIC} and their animated counterparts (see also Fig.~7 of \citealt{Hu22}). The more settled larger grains (e.g., the 10~mm grains) tend to concentrate more in the midplane fast accretion layer, particularly at relatively early times and large radii (see animation), where they are advected by the accreting gas radially inward more quickly. In contrast, the less settled smaller grains (e.g., the 0.1~mm grains) are affected more strongly by the meridional circulations at higher altitudes, where the gas circulates inward and outward and is thus less efficient in net inward grain advection. These differences between large and small grains can also be seen in the animated version of Fig.~\ref{fig: 2D_10mm01mm5bins_rings_gaps}. They are the reason why the largest (10~mm) grains are so strongly depleted relative to the gas compared to the smallest (0.1~mm) grains at relatively large radii between $\sim 10$ and $\sim 25$~au, especially in the gas gaps (see the blue curve in Fig.\ref{fig: 2D_10mm01mm5bins_rings_gaps}g near, e.g. $\sim 17$ and $\sim 23$~au). The depletion is less severe in the rings of the region (e.g., the one at $\sim 15$~au), likely because the higher density and stronger meridional circulations there can better prevent even the largest grains from concentrating in the fast accretion stream (and thus avoiding being rapidly advected inward). 

\begin{figure*}
    \centering
    \includegraphics[width=\linewidth]{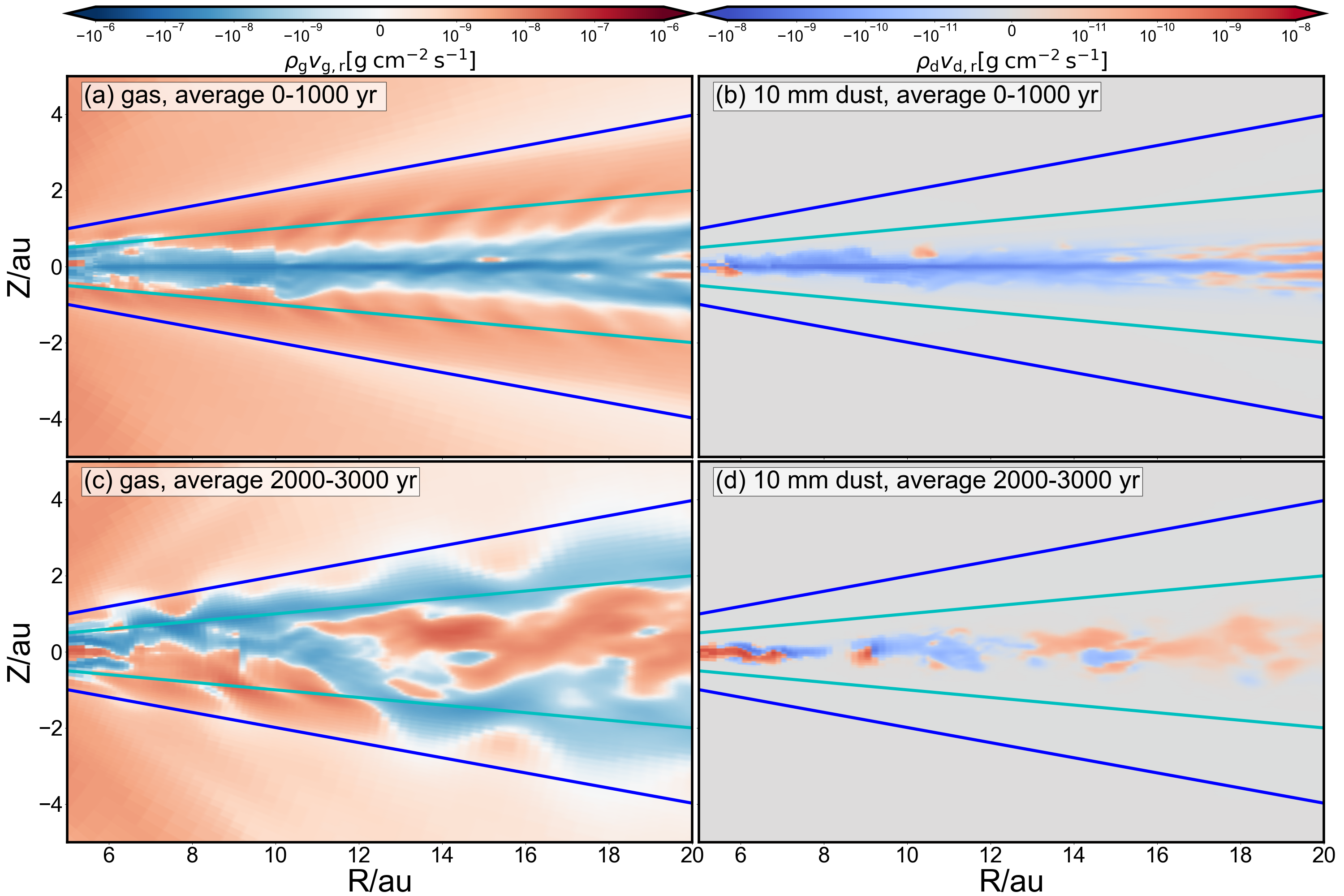}
    \caption{Distributions of the radial mass flux per unit area for the gas (left panels) and the largest (10~mm) grains (right panels) on a meridional plane averaged over the first 1000~yr (top panels) and the last 1000~yr (bottom panels). An animation of the time evolution of the instantaneous mass fluxes up to 3000~yr can be found at \url{https://figshare.com/s/1993dc8dd36c68063c0b}. 
    }
    \label{fig: MassFlux}
\end{figure*}

The large grains depleted from the outer $\sim 10-25$~au region are preferentially trapped in the inner region between $\sim 5-10$~au. The trapping appears to be primarily controlled by the vertical stratification of the gas accretion flow in the disk, which is more concentrated near the midplane at relatively early times. This is illustrated in panel (a) of Fig.~\ref{fig: MassFlux}, which plots the average radial mass flux per unit area of the gas in the first 1000~yr. During this time, the largest (10~mm) grains, which settle closest to the midplane, are advected predominantly inward radially by fast gas accretion streams near the midplane (see panel [b]). Later, the gas accretion streams move closer to the surface of the disk (see panel [c]), particularly inside $\sim 10$~au, where the gas accretion is concentrated on one side of the disk (the top surface). It is also the side where the poloidal magnetic field lines are the most pinched (see, e.g., Fig.~\ref{fig: LIC}a). The radially pinched field lines extract angular momentum from the surface accretion stream, one part of which is removed by the disk wind.  The remaining part of the extracted angular momentum is magnetically transferred to the disk material closer to the midplane below the stream, pushing it to expand outward, which slows or even reverses the inward migration of the thin dust layer near the midplane (see panel [d]). This surface accretion-driven midplane gas expansion, coupled with strong meridional circulations/vortices in gas rings \citep[e.g.][]{Hu22,Hsu24}, plays a key role in preferentially trapping dust in the inner 5-10~au region relative to the gas. A similar interplay between size-dependent dust settling, surface accretion-driven midplane gas expansion, and meridional vortices is also responsible for producing the strong enhancement of dust relative to gas at the outer $\sim30$~au ring (see Fig.~\ref{fig: 2D_10mm01mm5bins_rings_gaps}g).

\begin{figure}
    \centering
    \includegraphics[width=\linewidth]
    {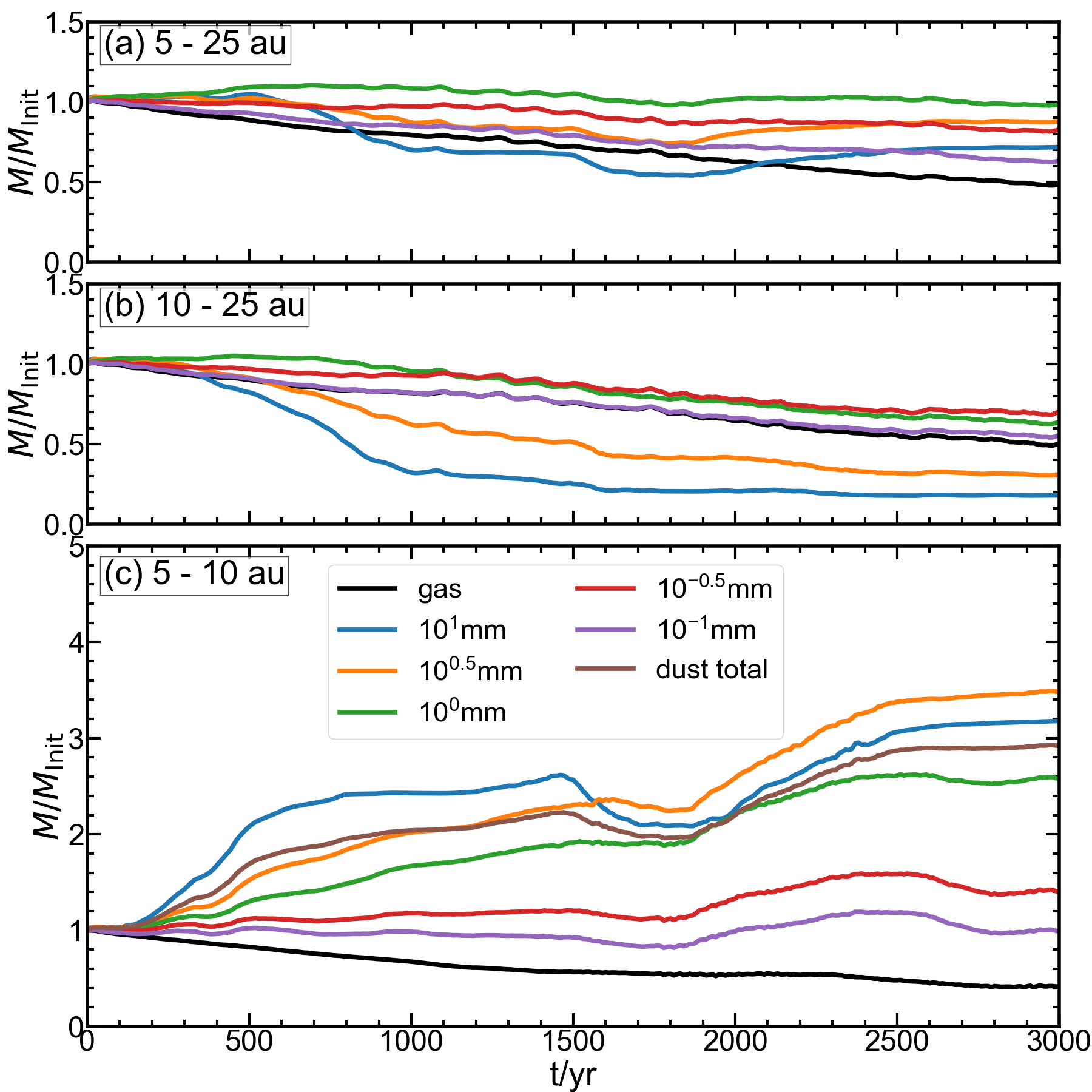}
    \caption{
    Evolution of total dust mass in three different regions (5-25 au [panel a], 10-25 au [b], and 5-10 au [c]), showing preferential depletion and concentration of dust relative to gas in the outer (10-25 au) and inner (5-10 au) disk regions, respectively. 
    }
    \label{fig: MassEvol}
\end{figure}

To further quantify the degree of radial redistribution in dust and gas, we plot in Fig.~\ref{fig: MassEvol} the masses of gas and grains of different sizes in 5-10~au, 10-25~au, and 5-25~au, normalized by their respective initial values. It is clear that for the 5-25~au region as a whole, there is constant depletion of gas, presumably through a combination of radial mass accretion through the disk and vertical gas removal by the disk wind, which has reduced the gas mass by $\sim 52\%$ at the end of the simulation at 3000~yr (see panel a). The dust is depleted to a lesser extent, and the degree of depletion depends on the grain size in a non-monotonic way. The smallest (0.1~mm) grains are depleted the most, as expected, since they are expected to follow the gas most closely. However, it is followed by the largest (10~mm) rather than the second smallest ($10^{-0.5}$~mm) grains. The least depleted dust in this region turns out to be the intermediate 1~mm-sized grains. 

The ordering of the degree of depletion is strongly region-dependent. In the outer (10-25~au) region, the largest grains are the most depleted, retaining only $\sim 18\%$ of their initial mass by the end of the simulation (see panel b). It is followed by the second largest ($10^{0.5}$~mm) grains, although the 1~mm grains are still less depleted than the smallest (0.1~mm) grains in this region. In contrast, in the inner (5-10~au) region, the masses of the three largest grains are strongly enhanced relative to their initial values, by factors up to $\sim 3.2$, $3.5$, and $2.6$, respectively, for the 10, $10^{0.5}$, and 1~mm grains by the end of the simulation (see panel c). The enhancement occurs despite the gas mass in the same region being depleted to only $\sim 40\%$ of its initial value. The net effect is a large increase in the dust-to-gas mass ratio in the region, reaching a factor of $\sim$7 at the end of the simulation. Such enhancement is expected to facilitate planetesimal formation, as we discuss in \S~\ref{sec: discussion} below. 

\begin{figure}
    \centering
    \includegraphics[width=\linewidth]
    {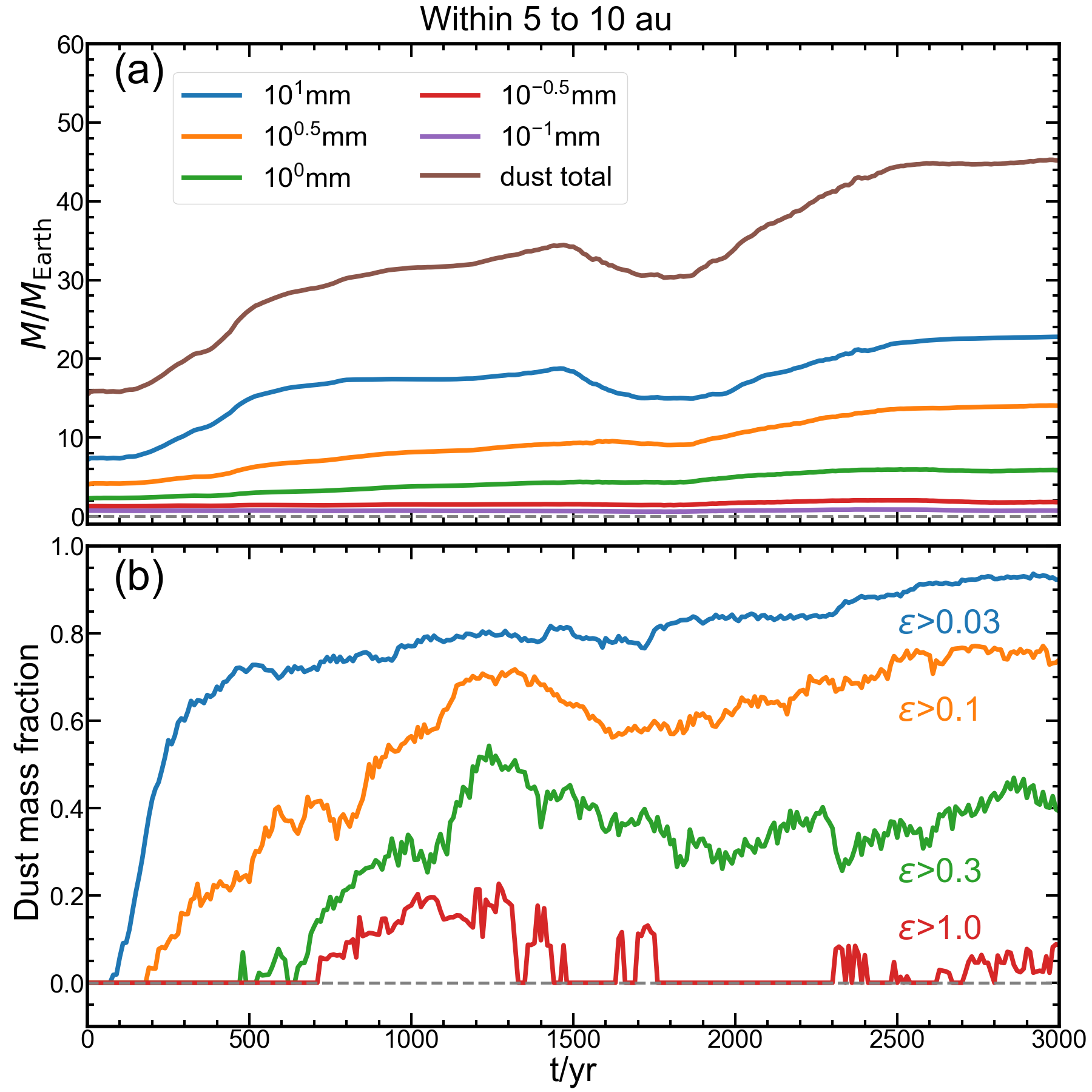}
    \caption{
    Dust concentration in the inner (5-10~au) region. Panel (a) shows the total dust mass in units of the Earth's mass as a function of time. Panel (b) shows the fraction of dust mass in regions above $0.03$, $0.1$, $0.3$, and $1.0$ times the local gas mass. 
    }
    \label{fig: DustInInnerRegion}
\end{figure}

To put the preferential dust concentration relative to the gas in the 5-10~au region into a broader context of planet formation, Fig.~\ref{fig: DustInInnerRegion}a shows the total dust mass in this region in units of Earth's mass as a function of time. It is clear that the dust mass in this inner disk region has increased by a factor of 3 by the end of the simulation, from $\sim 15$ to $\sim 45$~M$_\oplus$, enough to form the rocky cores of several Jupiter-like planets. Panel (b) shows that $\sim 2/3$ of the dust in this region is concentrated in regions with a dust-to-gas mass ratio $\epsilon > 0.1$ (that is, 10 times the initial value; see the orange curve) and $\sim 40\%$ (or $\sim$18~M$_\oplus$) in regions with $\epsilon > 0.3$ (the green curve). This concentration of dust in regions with greatly enhanced dust-to-gas mass ratios is expected to facilitate the formation of planetesimals through streaming instability, as we discuss in \S~\ref{sec: discussion} below. 

\subsection{Spatial variation of grain size distribution}
\label{subsec: GrainSizeDistribution}

\begin{figure}
    \centering
    \includegraphics[width=\linewidth]{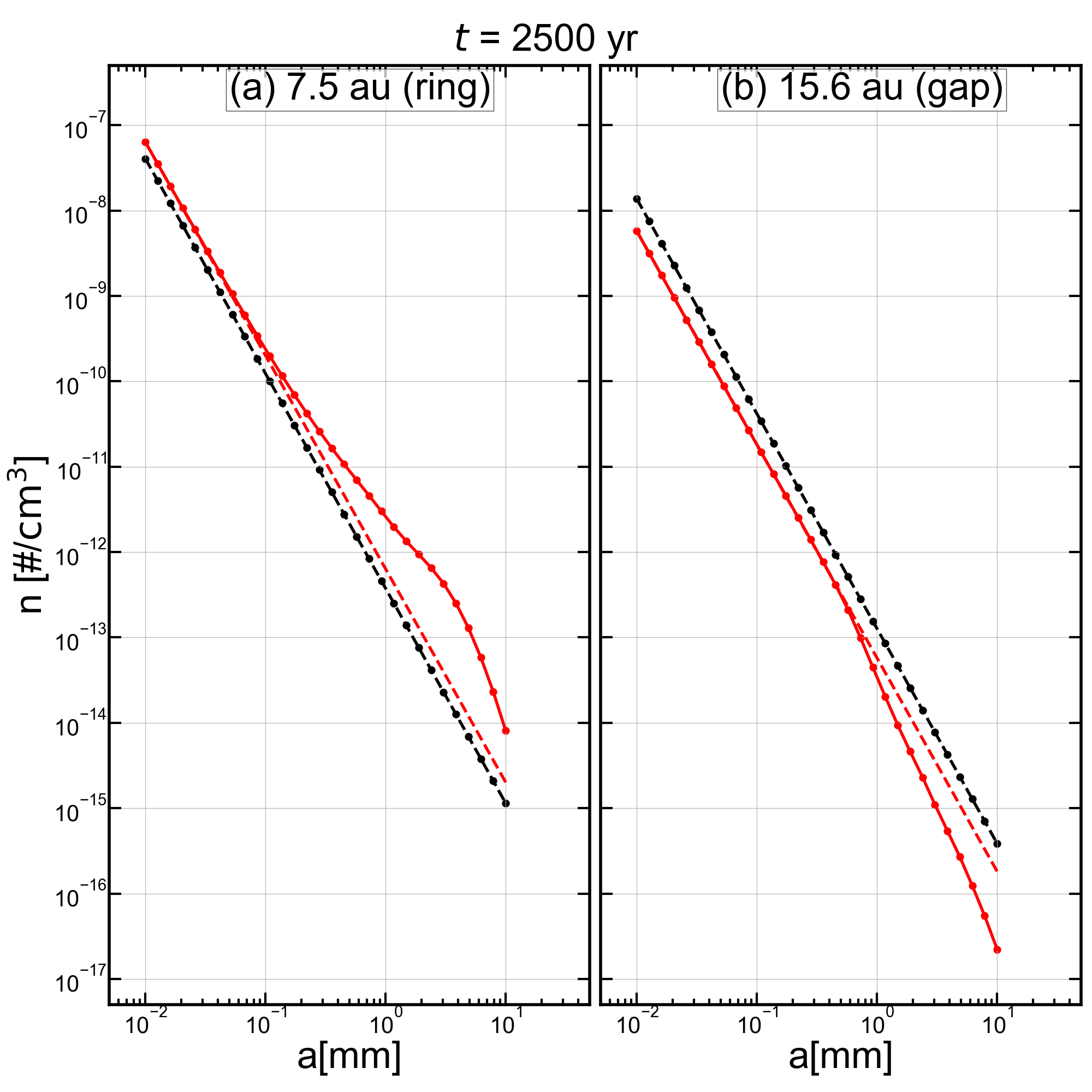}
    \caption{
Illustrative dust size distributions in a ring (panel a, $\sim$7.5~au) and a gap (panel b, $\sim$15.6~au) for the 30-dust-bin model 2D10mm10$\mu$m30bins. Black dots are the initial number density distribution.  Red dots are the number density distributions at $t=2500$~yr.  Red dashed lines are the lines with MRN distribution slope starting at the grain size $10^{-2}$~mm for reference.
    }
    \label{fig: size_distribution_30bins}
\end{figure}

The preferential concentration of larger grains relative to smaller grains in the 5-10~au region changes the distribution of dust size in the reference model 2D10mm01mm5bins. For example, at the location of the maximum dust-to-gas mass ratio for the largest grain at time $t=3000$~yr, the number of the largest grains increases by a factor of $\sim 55$ more than that of the smallest grains, leading to a flattening of the initial MRN-like power-law distribution. 
In contrast, in the outer 10-25~au region and away from the midplane, the larger grains are depleted more than the smaller grains, leading to a steeper dust size distribution. 

To illustrate the spatial variation of the dust size distribution more clearly, we repeated the 5-dust-size-bin simulation but with 30 bins and extended the minimum grain size from 0.1~mm (or 100~$\mu$m) to $10~\mu$m (Model 2D10mm10$\mu$m30bins), keeping the initial MRN power-law distribution. Rings and gaps are formed in both gas and dust, as in Model 2D10mm01mm5bins, although their positions are somewhat different, probably due to dust feedback on the gas dynamics, as we discuss in \S~\ref{subsec: dust_feedback} below. Nevertheless, the spatial variation of the dust size distribution remains, as illustrated in Fig.~\ref{fig: size_distribution_30bins}. Similarly to the 5-dust-bin case (discussed above but not plotted), the dust size distribution flattens for larger grains in the ring, creating a bump above the initial MRN distribution. At the time shown, the distribution can be approximated by three power laws, with the original MRN power law index of -3.5 for small grains between 10~$\mu$m to 0.2~mm, a shallower distribution with an index of -2.7 for the intermediate size range between 0.2~mm and 3~mm, and a steeper distribution with an index of -4.3 for the largest grains between 3~mm and 10~mm. In contrast, larger grains are depleted more with increasing size in the gap, with the power-law distribution steepening from an index of -3.5 at the small-size end to -4.4 at the large-size end. 

Spatial variations in dust density and size distribution may have implications on dust feedback on disk chemistry and dynamics, particularly the ionization level that controls the degree of magnetic coupling (see, e.g., \citealt{Dullemond18,Hu19}). 
They may also have observable effects. For example, larger grains in the rings can, in principle, emit and scatter longer wavelength photons more efficiently. Radiative transfer calculations based on appropriate opacities for the different grain size distributions in different regions are needed to quantify these effects. Such calculations and the feedback of the dust through its effect on the ionization level will be explored in future investigations.

\subsection{Dust aerodynamic feedback on gas substructure}
\label{subsec: dust_feedback}

\begin{figure}
   \centering
   \includegraphics[width=\linewidth]{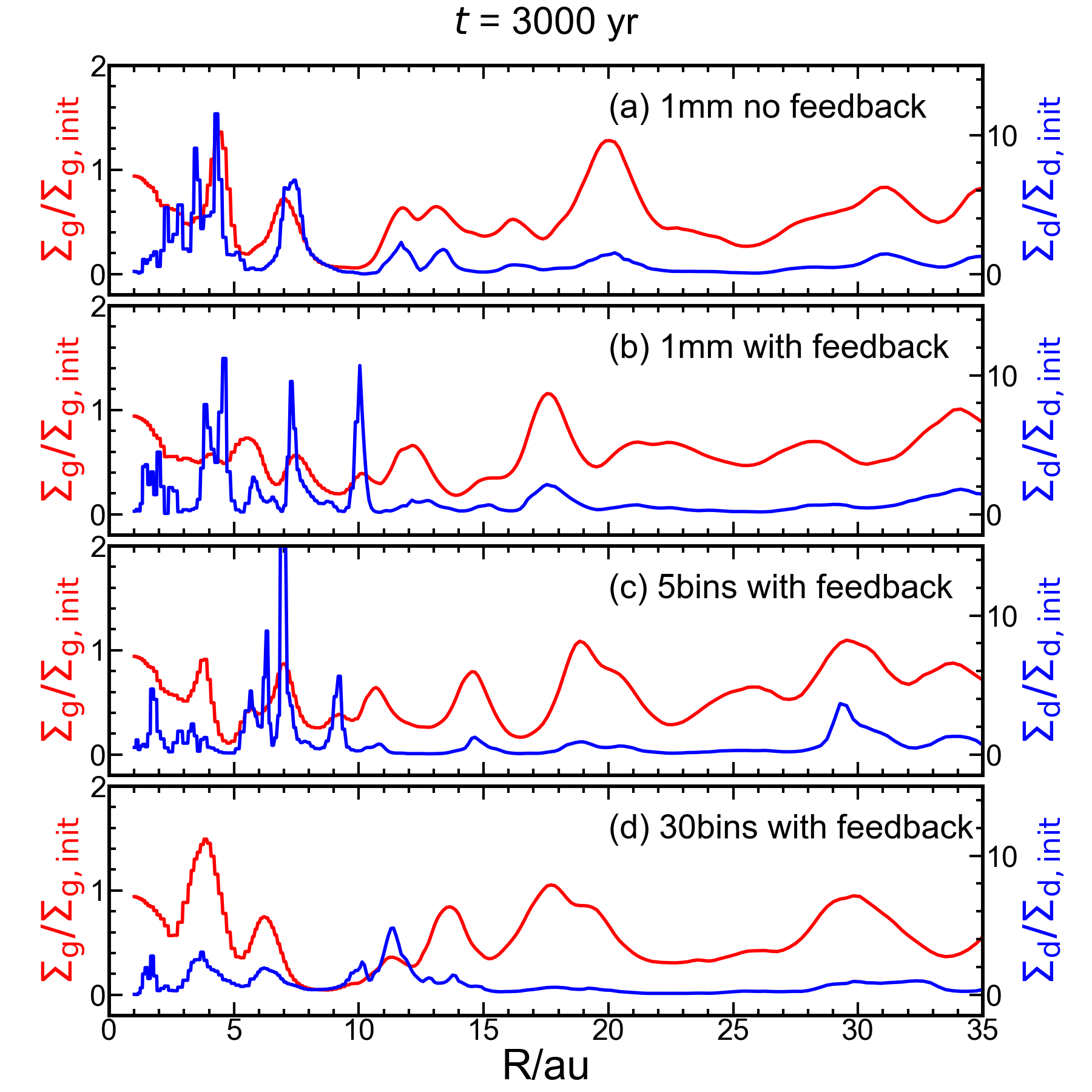}
    \caption{ 
    Comparison of the surface density distributions (normalized by their initial distributions) for the gas (red lines) and dust (blue lines) for all 4 2D (axisymmetric) models. Note the different plotting ranges for the normalized gas surface density (shown on the left vertical axis) and that for the dust (on the right), highlighting that dust tends to be concentrated more strongly than the gas. An animated version of the figure can be found on the website: \url{https://figshare.com/s/800c9797d357bbfa81fc}.
    }
    \label{fig: dust_feedback}
\end{figure}

\begin{figure*}
    \centering
    \includegraphics[width=\linewidth]{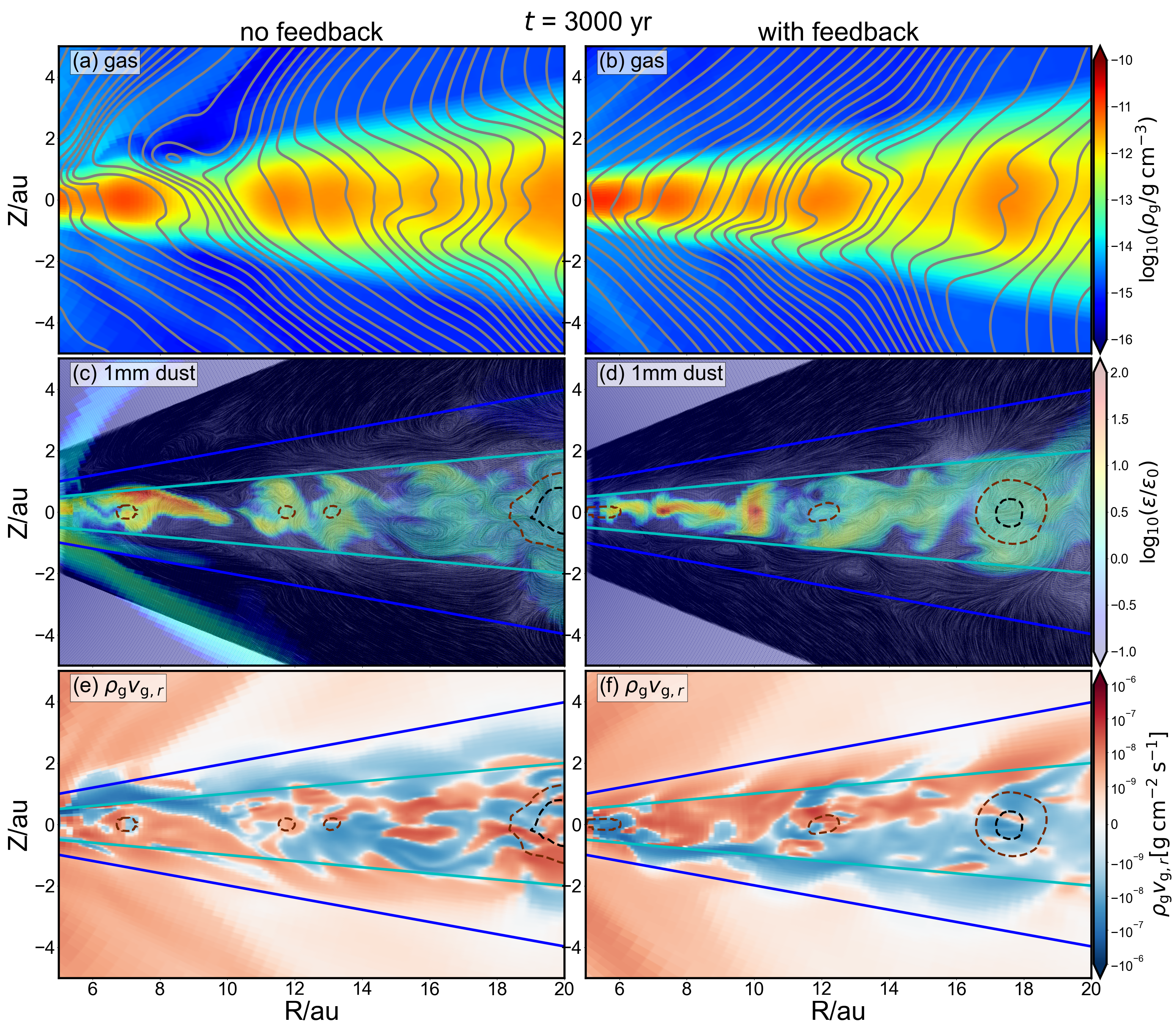}
    \caption{Dust feedback on the gas and dust substructure. Plotted in the left panels (a), (c), and (e) are, respectively, the gas density (color map) with poloidal field lines (gray) superposed, the mass density of 1~mm grains normalized by its initial value, with the LIC flow streamlines superposed, and gas mass flux per unit area in the radial direction ($\rho_{\rm g} v_{{\rm g}, r}$), for Model 2D1mm$\_$nofeedback. The right panels (b), (d), and (f) plot the same quantities but for Model 2D1mm. An animated version of the figure can be found on the website: \url{https://figshare.com/s/a0b88133eba193bc90b0}. 
    }
    \label{fig: dust_feedback_meridional}
\end{figure*}

Given the significant concentration of dust grains, particularly the largest ones, in localized regions of the disk that we discussed above, it is natural to examine the extent to which the dust affects the gas dynamics and, thus, its substructure formation through aerodynamic drag.  

Fig.~{\ref{fig: dust_feedback} and its associated animation compare the distributions of the gas and dust surface densities for the four 2D (axisymmetric) models listed in Table~\ref{table_cases}, including 2D1mm$\_$nofeedback, 2D1mm, 2D10mm01mm5bins, and 2D10mm10$\mu$m30bins. From a comparison of the first two models (panels [a] and [b]), it is apparent that the dust feedback in the latter model did not suppress the formation of rings and gaps in the gas and dust, but has substantially modified their properties, including their locations and amplitudes. For example, there is a moderate gas ring and a strong dust ring around 10~au at the time shown with dust feedback (panel [b]), but only a gap in both gas and dust at the same location without feedback (panel [a]). The contrast can be seen more pictorially in Fig.~\ref{fig: dust_feedback_meridional}, which plots the gas and dust distributions on the meridional plane, together with the poloidal magnetic field lines (top panels) and flow streamlines (middle panels). 

The exact reason for the differences is unclear. However, they are not unexpected given the rather chaotic and highly dynamic nature of the gas motions in the bulk of the disk and near the base of the disk wind (as illustrated by the streamlines in Fig.\ref{fig: dust_feedback_meridional}[b] and [d] and the gas mass flux in the radial direction in Fig.\ref{fig: dust_feedback_meridional}[e] and [f]). Even small changes in gas motion induced by dust aerodynamic drag can accumulate over time, leading to significant differences in the gas (and dust) substructures through non-linear interactions between the gas and magnetic field inside both the disk and the disk wind, and between these two regions. In particular, radially pinched poloidal field lines conducive to strong field twisting (and thus magnetic braking) often develop in the transition zone between the disk and wind, leading to strong surface accretion, as discussed in the last subsection (in connection with dust concentration). The pinched field and associated surface accretion tend to be more prominent on one side of the disk than on the other at a given time \citep[see, e.g.,][]{Bai14,Bethune17,Riols19}, but which side is stronger depends on the rather chaotic evolution of the toroidal magnetic field inside the disk and is difficult to predict. Since surface accretion with radially pinched poloidal field lines is a form of MRI-unstable channel flow \citep[e.g.,][]{Balbus02}, it can run away and produce highly pinched field lines prone to reconnection \citep[see, e.g.,][]{Tu25}. An example of this process is shown in panels (a) and (e) of Fig.~\ref{fig: dust_feedback_meridional}, where a strong surface accretion occurs on the upper side inside $\sim$10~au at the time shown, with a closed magnetic loop from reconnection; the strong magnetic braking from the highly pinched field lines is probably the reason behind the gas and dust gaps around $\sim$9~au shown in Fig.~\ref{fig: dust_feedback}. In contrast, the dust feedback in Model 2D1mm appears to have moved the surface accretion to the lower side of the disk (see panel [f]), and the field lines in the region are still radially pinched but not strong enough to cause visible reconnection (see panel [b]). In any case, the aerodynamic feedback of the dust clearly affects the substructures of the gas and dust, but the exact mechanism appears complex and remains uncertain. 

The dust feedback depends on grain size distribution, as illustrated in the two lowest panels of Fig.~\ref{fig: dust_feedback}. For example, there is a prominent gas ring at $\sim 14.5$~au in Model 2D10mm01mm5bins (panel [c] and Fig.~\ref{fig: 2D_10mm01mm5bins_rings_gaps}a) that is absent from Model 2D1mm (panel [b] and Fig.~\ref{fig: dust_feedback_meridional}b), indicating that the presence of grains of different sizes, particularly larger (e.g., 10~mm) grains, in the former affected the gas dynamics differently than the single-sized (1~mm) grains. The gas flow structures are also different, with the pinched poloidal field (Fig.~\ref{fig: LIC}a) and its associated fast accretion concentrating on the upper side of the disk in the 5-dust-bin model but on the lower side in the single-size grain case (Figs.~\ref{fig: dust_feedback_meridional}[b] and [f]). In addition, there are visible differences between the two multi-size-bin models 2D10mm10$\mu$m30bins and 2D10mm01mm5bins, particularly in the region between $\sim 5$-$10$~au at the time shown in Fig.~\ref{fig: dust_feedback}, where gas rings have different locations, and the dust is much more concentrated in the latter than the former. These differences are caused, at least in part, by strong accretion developing near both the upper and lower surfaces in the 30 dust-bin model around the time shown, with strongly pinched poloidal field lines that reconnect to produce magnetic loops on both sides (not plotted). They indicate that the gas distribution and the dynamics of the disk are affected by the lower limit ($a_{\rm min}$) of the grain size distribution and possibly the number of size bins, which are 0.01~mm and 30, respectively, for the former and 0.1~mm and 5 for the latter. In any case, the dust aerodynamic feedback does not change the qualitative conclusion that both rings and gaps are spontaneously formed in the gas and dust but does modify the properties of the formed substructures quantitatively. The complex interaction between dust aerodynamic feedback, gas dynamics, magnetic fields, and magnetized disk wind warrants further detailed exploration in the future.

\subsection{Dust concentration in 3D} 
\label{subsec: 3D_single_fluid}

\begin{figure*}
    \centering
    \includegraphics[width=\linewidth]{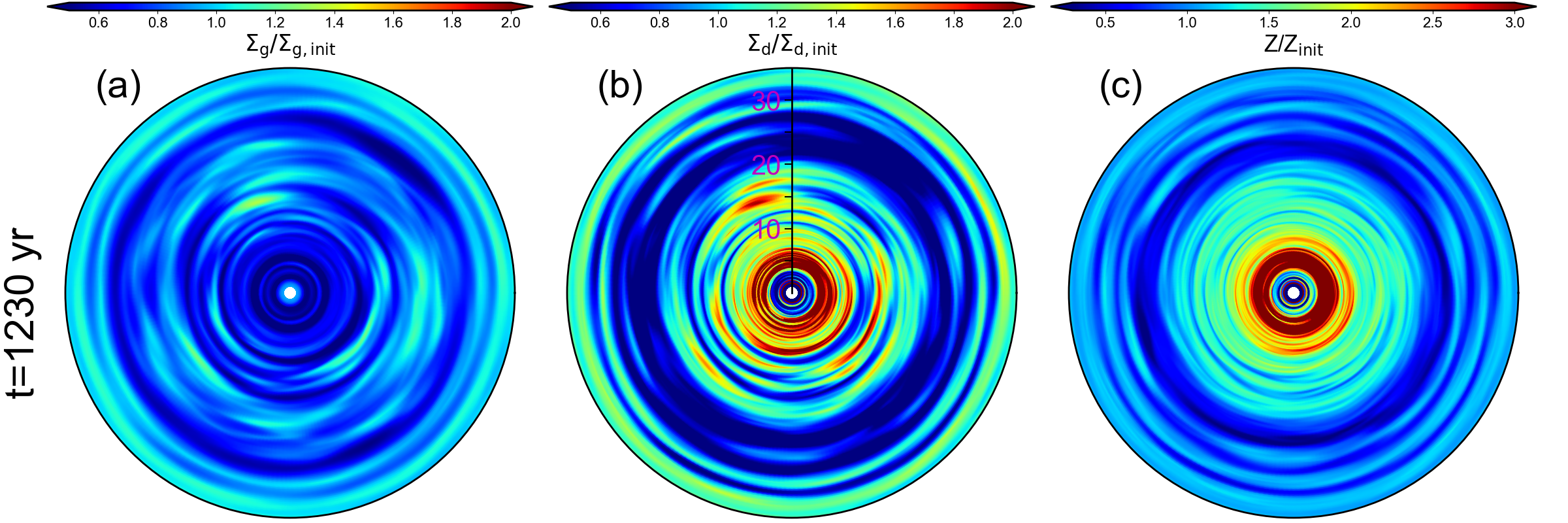}
    \caption{
    Non-axisymmetric substructures in gas and dust in Model 3D1mm.  Plotted are the surface densities of the gas (panel [a]) and dust (panel [b]) normalized by their respective initial values and dust-to-gas surface density ratio or metalicity $Z$ (normalized by the initial value of 0.01; panel [c]). An animated version of the figure can be found at \url{https://figshare.com/s/2d2c8126ece57dbbc094}.
    }
    \label{fig: 3D_dust_gas_contour}
\end{figure*}

\begin{figure*}
    \centering
    \includegraphics[width=\linewidth]{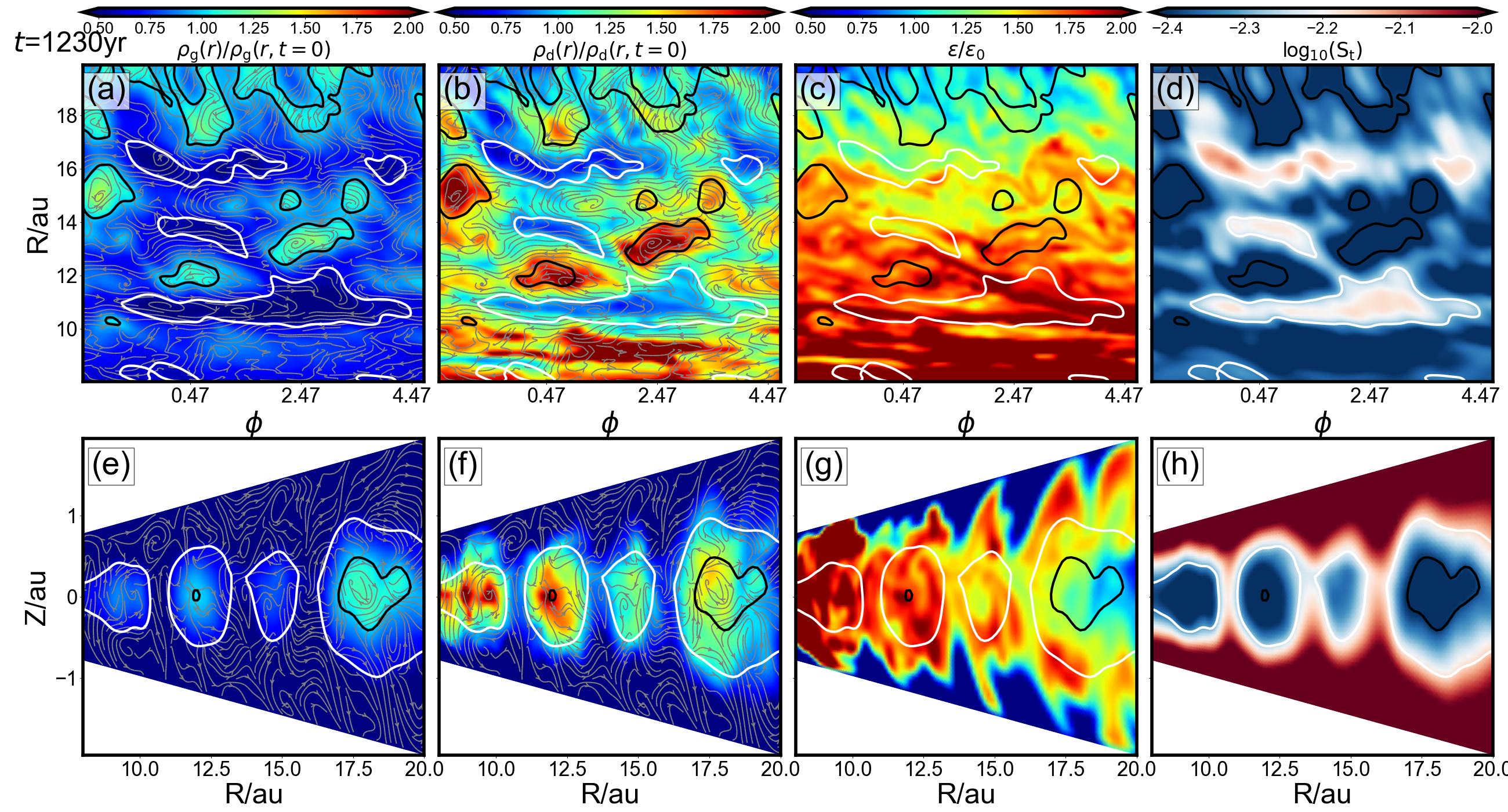}
    \caption{
    Dust concentration in 3D Model 3D1mm. Plotted in the upper panels are the midplane $R$-$\phi$ distributions of the gas density (panel [a]), dust density (panel [b]), the dust-to-gas mass ratio (panel [c]), normalized by their respective initial values, and the Stokes number of the dust (panel [d]), at a representative time $t=1230$~yr. The corresponding distributions of the same quantities on a meridional ($R$-$z$) plane are shown in panels (e)-(h). The gray lines in the left half of the figure (panels [a], [b], [e], and [f]) are velocity streamlines, showing the vortices. Also plotted in each panel are contours marking where the gas density is, respectively, 0.6 (white) and 1.0 (black) times its initial value. Note that the white contours highlight the gaps in the upper panels but the rings in the lower ones. An animated version of the figure can be found at \url{https://figshare.com/s/8efc3c47141d274ed7ab}. 
    }
    \label{fig: 3D_vortice_parameters}
\end{figure*}

In addition to the 2D (axisymmetric) simulations discussed in the preceding subsections, we have performed a 3D simulation with single-sized (1~mm) grains (Model 3D1mm in Table~\ref{table_cases}). It can be viewed as an extension of \cite{Hsu24}, who investigated the 3D evolution of the gas substructures formed in non-ideal MHD wind-launching discs but without dust. Our gas plus dust simulation shows a gas distribution and evolution broadly consistent with the gas-only case, with non-axisymmetric features developing on rings and gaps from vortices produced by RWI \citep[Rossby wave instability, e.g.][]{Lovelace99}. These features are illustrated in Figs.~\ref{fig: 3D_dust_gas_contour} and \ref{fig: 3D_vortice_parameters} and their associated animations. Comparing panels (a) and (b) of Fig.~\ref{fig: 3D_dust_gas_contour}, it is clear that arc-shaped nonaxisymmetric substructures are formed in both gas and dust, particularly around $\sim 15$~au at the representative time shown (t=1230~yr). However, the dust arcs stand out more than their gas counterparts in the plots. This is because the mm-sized grains have preferentially migrated radially inward from larger to smaller radii relative to the gas, leaving deeper dust gaps with dust-to-gas mass ratios below the initial value between $\sim 20$ and $\sim 30$~au (see panel [c]) and preferential concentration of dust relative to gas in the rings at smaller radii, similar to the 2D (axisymmetric) cases (see, e.g., Fig.~\ref{fig:2D_1mm_rings_gaps}); the preferential dust concentration makes the arcs more prominent in the surface density distribution normalized to its initial value. 

Fig.~\ref{fig: 3D_vortice_parameters} and its associated animation show that the arc-like surface density features are vortices. Specifically, panel (a) shows that the flow (after subtracting the Keplerian rotation) on the midplane tends to circulate around the denser parts of the rings (highlighted by black contours) in the clockwise direction and around the lower-density parts of the gaps (highlighted by white contours) in the counterclockwise direction, as expected for RWI-generated vortices and discussed in depth in \cite{Hsu24}. Panel (b) shows that, similar to the gas, the mm-sized dust is enhanced in the same clockwise circulating vortices and depleted in the counterclockwise ones. There is evidence that some denser regions trap the mm-sized grains preferentially relative to the gas than lower-density regions at the same radius (compare the two elongated regions enclosed by the black and white contours near $\sim 13.5$~au in panel [c], where the dust-to-gas mass ratio is plotted), but there is no one-to-one correspondence between regions of gas density enhancement and dust-to-gas mass ratio enhancement. In other words, denser regions at a given radius do not always preferentially trap more dust relative to the gas than lower-density regions at the same radius. This is likely a result of the highly dynamic nature of the disk substructures in our simulation and the mm-sized grains being coupled to the gas, with a Stokes number ${\rm St}$ of order $10^{-2}$ or less on the midplane in the region shown in panel (d). Note that the Stokes number is higher in lower-density regions, indicating that the grains are less coupled to the gas there than in denser regions, as expected. 

The highly dynamic nature of the gas flow is also evident from panel (e) of Fig.~\ref{fig: 3D_vortice_parameters} and its animated version, which shows vigorous meridional circulation motions that were examined in detail in \cite{Hu22} and \cite{Hsu24}. The interplay between these motions and the vertical settling and radial migration of dust complicates the dynamics of the dust. Nevertheless, it remains true that the dust distribution broadly follows that of the gas (comparing panels [e] and [f]), with a general tendency of preferential dust concentration relative to the gas at smaller radii, but no one-to-one correspondence between the dust-to-gas mass ratio and gas density enhancement (see panel [g]). Panel (h) reinforces the result that the mm-sized grains have the lowest Stokes number in the gas rings and are thus best coupled to the denser gas there. This makes it easier to keep the dust entering the rings from the lower-density surrounding regions, such as the gas gaps and the regions above and below the rings, where the dust is less well coupled to the gas. 

\section{Potential early planetesimal formation in actively accreting disks}
\label{sec: discussion}

\begin{figure*}
   \centering
   \includegraphics[width=\linewidth]{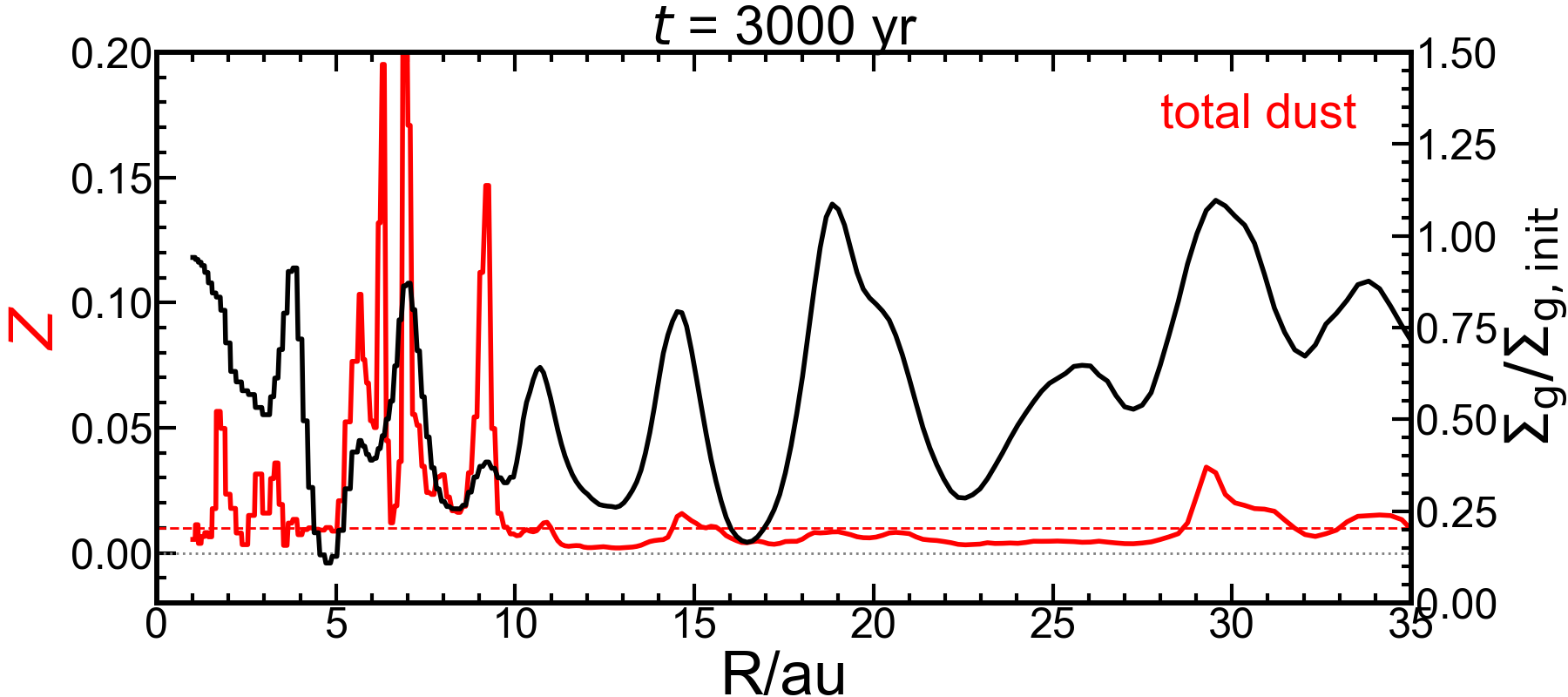}
    \caption{Distribution of the dust-to-gas mass surface density ratio (a.k.a metalicity $Z$) as a function of radius for the 2D10mm01mm5bins model, showing values well above the initial value of 0.01 (marked by the horizontal red dashed line), particularly between 5 and 10~au. An animated version of the figure can be found at \url{https://figshare.com/s/f4a11b000e21decc1963}.  
    }
   \label{fig: Z}
\end{figure*}

\begin{figure}
   \centering
   \includegraphics[width=1.0\linewidth]{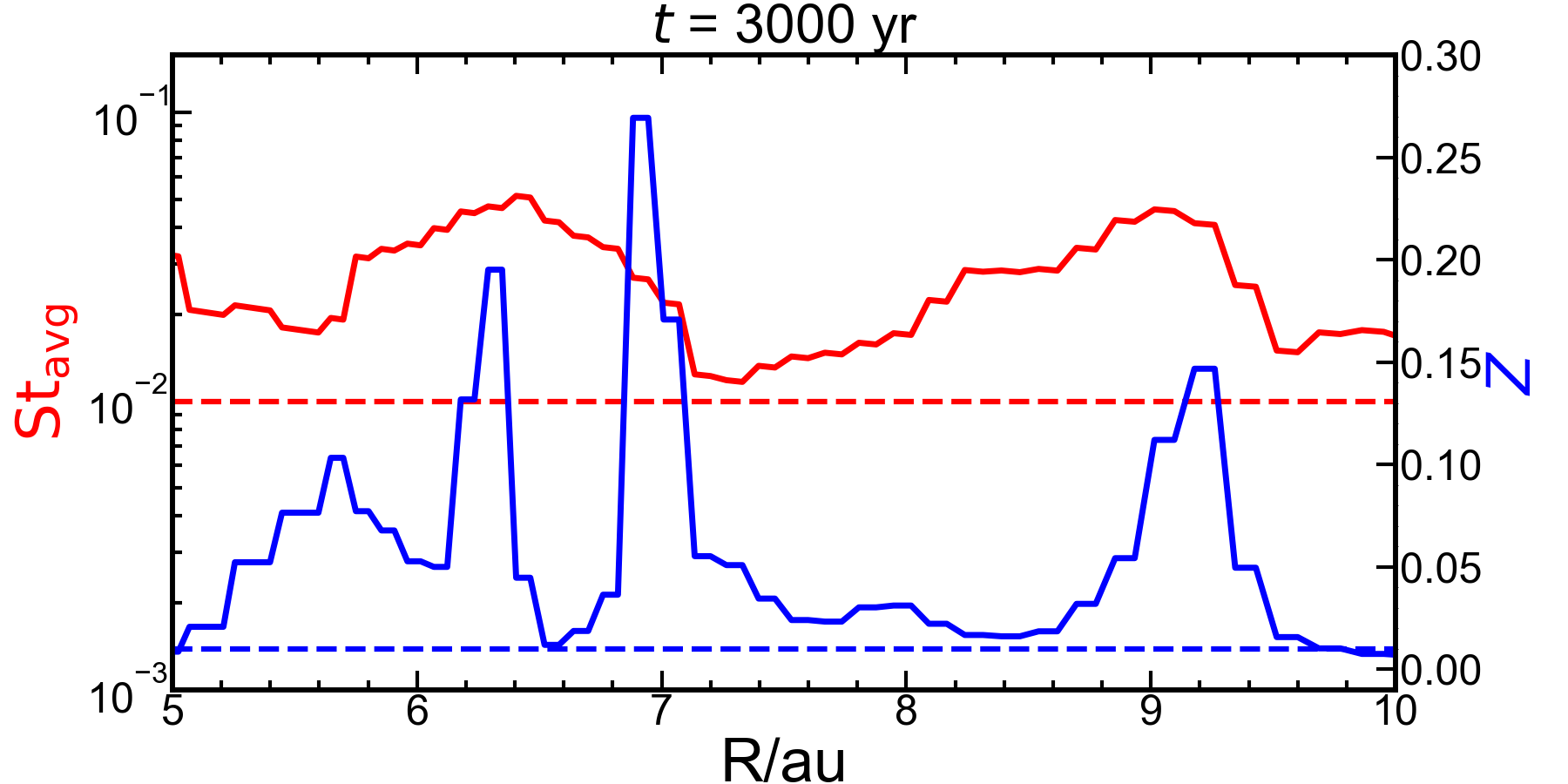}
    \caption{The distribution of the average dust Stokes number $\rm{St}$ as a function of radius in the inner (5-10~au) disk region (red curve) where the dust-to-gas mass surface density ratio $Z$ (blue curve) is greatly enhanced above its initial value of 0.01 (the blue dashed line) at the same time as in Fig.~\ref{fig: Z}. The red horizontal dashed line marks $\rm {St}=10^{-2}$. 
    }
   \label{fig: St}
\end{figure}

One of the most important results of our work is that mm/cm-sized grains can concentrate strongly on the 5-10 au scale (see Fig.~\ref{fig: MassEvol}c), with $\sim 40\%$ of the dust residing in localized regions where the dust-to-gas mass ratio $\epsilon > 0.3$ (see Fig.~\ref{fig: DustInInnerRegion}, lower panel), despite vigorous meridional circulation motions and strong accretion driven by the removal of angular momentum via magnetic disk wind. The high degree of dust concentration in an actively accreting disk opens up the possibility of early planetesimal formation during the Class 0 and Class I phases of star formation if grains have already grown to such relatively large sizes, as indicated by the frequent detection of embedded disks at centimeter wavelengths \citep[e.g.,][]{Segura16}. 

The already high dust concentration in our global simulations can be further enhanced by the streaming instability (SI), which is explored primarily through linear stability analysis \citep[e.g.,][]{Youdin05} and local shearing box simulations \citep[e.g.,][]{Johansen07,Johansen14}. 
Although our global non-ideal MHD simulations do not have the resolution to resolve SI, they do provide estimates for the global gas and dust quantities that control the behavior of dust clumping in local shearing box simulations, including the vertically integrated dust-to-gas mass ratio $Z$ (a.k.a. ``metalicity"), the dust Stokes number ${\rm St}$, the Toomre parameter $Q$, and the radial velocity of the disk material $v_r$. 

In a typical vertically stratified shearing box simulation of SI in a quiescent laminar background, whether dust clumping is strong enough to reach the local Roche density to potentially gravitationally collapse and form planetesimals is primarily controlled by metalicity $Z$ and dust Stokes number ${\rm St}$ \citep[e.g.,][]{Johansen14}. For example, \cite{Rixin21} found critical values for $Z$ for strong clumping of $\sim 0.007$ (or lower) for relatively large grains with ${\rm St} \gtrsim 10^{-2}$ (reaching values as low as 0.004 for ${\rm St} \sim 0.3$; see their Fig.~1). As ${\rm St}$ decreases from $\sim 10^{-2}$ to $\sim 10^{-3}$, the critical value increases from $\sim 0.02$ to $\sim 0.04$, higher than the canonical value of $0.01$, which is also the initial value of our global simulation. The preferential concentration of dust relative to gas in our simulations increases the dust-to-gas surface density ratio $Z$ well beyond 0.01 in localized regions, as illustrated in Fig.~\ref{fig: Z} and its animated version for the reference model 2D10mm01mm5bins discussed in detail in \S~\ref{subsec: 2D_muti-fluid}. The high $Z$ values are conducive to dust clumping through SI. 

 The development of SI in the high $Z$ regions between 5 and 10 au is expected to be aided by a relatively large average Stokes number ${\rm St}$, which ranges from $\sim 0.01$ to $\sim 0.05$ at the representative time shown in Fig.~\ref{fig: St} (see panel [a]); as mentioned earlier, such values of ${\rm St}$ are favorable for SI, with a corresponding critical $Z$ for strong clumping ranging from $\sim 0.004$ to $\sim 0.015$ in the set of simulations performed by \cite{Rixin21}. 
It is potentially further enhanced by two characteristics of the Class 0 and I disks. Firstly, such young disks are expected to accrete at higher rates, possibly due to magnetic torques (e.g., \citealt{Tu24}; see, however, \citealt{Xu21}, for a different view). In such cases, a variant of the SI -- the so-called ``azimuthal-drift" streaming instability (AdSI) -- can significantly lower the threshold for the dust-to-gas mass ratio for substantial dust clumping, particularly near gas pressure maxima where large grains tend to be trapped \citep[][]{Lin22,Hsu22,Wang24}. Secondly, the youngest Class 0 and I disks are expected to be closer to the gravitational instability than the more evolved Class II disks because of lower stellar masses and possibly higher disk masses, with a Toomre parameter $Q$ not too far from unity (see, e.g., \citealt{Xu21}), even when the disk accretion is dominated by magnetic braking (see, e.g., \citealt{Tu24}). For example, \cite{LinZY21} found indirect evidence for a relatively low $Q$ value in the prototypical, edge-on, Class 0 disk HH212: in order for the dust opacity at 1.3~mm derived from the spatially resolved dust continuum intensity profile along the disk's major axis to agree with the commonly used opacity of \cite{Beckwith90}, we must have $Q\sim 2$ (see their Fig.~11), which is much lower than the value of $Q=32$ adopted in \cite{Rixin21}. Since the local dust density $\rho_{\rm R}$ needed to reach the Roche density for gravitational collapse relative to the midplane gas density $\rho_{\rm g0}$ is linearly proportional to $Q$ ($\rho_{\rm R}/\rho_{\rm g0} = 5.6~Q$; see equation~[5] of \cite{Rixin21}), the much smaller $Q$ for the youngest disks may greatly lower the dust density threshold (relative to the gas) for gravitational collapse to form planetesimals. 

The dust clumping may, however, be limited by gas motions in the actively accreting disk, particularly the meridional circulation that can be clearly seen in the streamlines plotted in Figs.~\ref{fig: LIC} and \ref{fig: 3D_vortice_parameters} and their respective animations. It is well known that even a relatively low level of background (isotropic, microscopic) turbulence can adversely affect the SI by diffusing small-scale motions and dust density variations (see, e.g., \citealt{Lesur22}). However, it remains unclear to what extent the relatively large-scale meridional circulation acts as isotropic turbulence, particularly on the small scale of the thin dust layer. Future investigations that combine large-scale global simulations with small-scale local ones are needed to directly address this question. 
%
%

\section{Conclusion} \label{sec: conclusion}

We explored the dynamics of dust concentration in actively accreting, substructured, non-ideal MHD wind-launching disks through 2D and 3D simulations. These simulations incorporate pressureless dust fluids of different grain sizes and account for their aerodynamic feedback on gas dynamics. Our main findings are as follows.

\begin{enumerate}
 \item We find that mm/cm-sized grains are preferentially concentrated in localized zones of the disk, particularly within the inner 5–10 au, where the dust-to-gas surface density ratio (commonly referred to as metallicity Z) significantly exceeds the canonical value of 0.01. This preferential dust concentration arises from the interplay between dust settling and complex gas flow patterns in the non-ideal MHD wind-launching disk. In particular, well-settled large grains are initially advected inward by fast midplane accretion streams to small radii, where they are subsequently trapped dynamically by the midplane gas expansion driven by angular momentum transfer from magnetically braked surface accretion streams and the meridional circulation in spontaneously formed gas rings. 
 The trapped dust is concentrated in localized pockets with high dust-to-gas mass ratios ($\epsilon$), with $\sim 40\%$ of the dust residing in regions of $\epsilon > 0.3$ in the 5-10~au region of our reference simulation at later times.

\item The combined effects of vertical settling and radial migration lead to a highly spatially variable grain size distribution. In particular, low-density gas gaps are more depleted in large grains, resulting in a steeper size distribution at the large-grain end compared to dense gas rings, where large grains preferentially accumulate. This variation in grain size is expected to influence the chemistry of the disk and ionization levels, which may affect the dynamics of the magnetized disk indirectly through the gas-field coupling. Direct aerodynamic feedback of dust on gas
modifies the quantitative properties of the rings and gaps but does not fundamentally alter their spontaneous formation in both gas and dust.

\item 
The rings and gaps formed in the 3D non-ideal MHD wind-launching disk simulation including dust aerodynamic feedback are unstable to Rossby wave instability (RWI). This leads to the formation of arc-shaped vortices in both gas and dust, similar to previous simulations without dust feedback. However, in the inner disk, where mm-sized grains are more concentrated relative to the gas, these dust arcs stand out more prominently than their gas counterparts.

\item The substantial enhancement of the local dust-to-gas surface density ratio (Z) beyond the canonical 0.01 is expected to facilitate dust clumping via the streaming instability, increasing the likelihood of early planetesimal formation in an actively accreting disk. This process may be further aided by the ``azimuthal-drift" streaming instability (AdSI), which operates more efficiently in strongly accreting disks and significantly lowers the dust-to-gas mass ratio threshold for substantial dust clumping. Additionally, the lower Toomre $Q$ parameter expected of younger disks reduces the dust density threshold (relative to the gas) required for gravitational collapse into planetesimals. More research is needed to confirm or rule out this promising possibility.
\end{enumerate}

\section*{Acknowledgements}
We thank the anonymous referee for a prompt and constructive
report.
We thank Pinghui Huang for his help with the dust module implemented in ATHENA++ that was used in this work. CYH acknowledges support from the NRAO ALMA Student Observing Support (SOS) and computing resources from UVA research computing (RIVANNA), NASA High-Performance Computing, and NSF ACCESS (AST200032, PHY240192). ZYL is supported in part by NASA 80NSSC20K0533, NSF AST-2307199, and the Virginia Institute of Theoretical Astronomy (VITA). 

\section*{Data Availability}
The data underlying this article will be shared on reasonable request to the corresponding author.


\bibliographystyle{mnras}
\bibliography{main} 





\bsp	
\label{lastpage}
\end{document}